\documentclass[preprint]{aastex}

\newcommand{\be}{\begin{equation}}
\newcommand{\ee}{\end{equation}}
\newcommand{\cdelt}{{ C_\Delta }}

\newcommand{\tmod}{{ [t] }} 
\newcommand{\af}{{ \lambda }} 
  
\newcommand{\lamwig}{{ {\widetilde \lambda}_j }} 
\newcommand{\qwig}{{ {\widetilde q}_j }} 
\newcommand{\grow}{{ f_{+} }} 
\newcommand{\decay}{{ f_{-} }}
\newcommand{\rat}{{ {\cal R} }}   
\newcommand{\qhat}{ {\hat Q} } 
\newcommand{\discrim}{ { \Delta } } 
\newcommand{\sumtn}{{ \Sigma_{T(N)} }} 
\newcommand{\sumbn}{{ \Sigma_{B(N)} }} 
\newcommand{\swig}{{ {\widetilde S} }} 
\newcommand{\dgamz}{{ \Delta \gamma_0 }} 
\newcommand{\podd}{{P_{\rm odd} }} 
\newcommand{\peven}{{P_{\rm even} }} 
\newcommand{\gem}{{ \gamma_{\rm em} }}
\newcommand{\czero}{{ C_0 }} 
\newcommand{\cbig}{{ C_\infty }} 
\newcommand{\thmbox}{{ \square }} 
\newcommand{\coneps}{{ K_\varepsilon}}
\newcommand{\fsubk}{{ \phi_k }}  
\newcommand{\fsub}{{ \phi }}  

\begin{document} 

\title{\bf HILL'S EQUATION WITH RANDOM FORCING TERMS} 

\author{Fred C. Adams$^{1,2}$ and Anthony M. Bloch$^{1,3}$} 

\affil{$^1$Michigan Center for Theoretical Physics \\
Physics Department, University of Michigan, Ann Arbor, MI 48109} 

\affil{$^2$Astronomy Department, University of Michigan, Ann Arbor, MI 48109} 

\affil{$^3$Department of Mathematics, University of Michigan, Ann Arbor, MI 48109} 

\begin{abstract} 

Motivated by a class of orbit problems in astrophysics, this paper
considers solutions to Hill's equation with forcing strength
parameters that vary from cycle to cycle. The results are generalized
to include period variations from cycle to cycle.  The development of
the solutions to the differential equation is governed by a discrete
map.  For the general case of Hill's equation in the unstable limit,
we consider separately the case of purely positive matrix elements and
those with mixed signs; we then find exact expressions, bounds, and
estimates for the growth rates. We also find exact
expressions, estimates, and bounds for the infinite products of
several $2 \times 2$ matrices with random variables in the matrix
elements.  In the limit of sharply spiked forcing terms (the delta
function limit), we find analytic solutions for each cycle and for the
discrete map that matches solutions from cycle to cycle; for this case
we find the growth rates and the condition for instability in the
limit of large forcing strength, as well as the widths of the
stable/unstable zones.

\end{abstract} 

\section{INTRODUCTION}  

This paper presents new results concerning Hill's equation of the form 
\be 
{d^2 y \over dt^2} + [ \af_k + q_k \qhat (t) ] y = 0 \, , 
\label{eq:basic} 
\ee 
where the function $\qhat (t)$ is periodic, so that $\qhat (t + \pi) =
\qhat(t)$, and normalized, so that $\int_0^\pi \qhat dt$ = 1. The
parameter $q_k$ is denoted here as the forcing strength, which we
consider to be a random variable that takes on a new value every cycle
(the index $k$ determines the cycle). The parameter $\af_k$, which
determines the oscillation frequency in the absence of forcing, also
varies from cycle to cycle. In principal, the duration of the cycle
could also vary; our first result (see Theorem 1) shows that this
generalized case can be reduced to the problem of equation
(\ref{eq:basic}).

Hill's equations [HI] arise in a wide variety of contexts [MW] and
hence the consideration of random variations in the parameters ($q_k,
\af_k$) is a natural generalization of previous work. This particular
form of Hill's equation was motivated by a class of orbit problems in
astrophysics [AB]. In many astrophysical systems, orbits take place in
extended mass distributions with triaxial forms. Examples include dark
matter halos that envelop galaxies and galaxy clusters, stellar bulges
found at the centers of spiral galaxies, elliptical galaxies, and
young embedded star clusters. These systems thus occur over an
enormous range of scales, spanning factors of millions in size and
factors of trillions in mass. Nonetheless, the basic form of the
potential is similar [NF, BE, AB] for all of these systems, and the
corresponding orbit problem represents a sizable fraction of the
orbital motion that takes place in our universe. In this context, when
a test particle (e.g., a star or a dark matter particle) orbits within
the triaxial potential, motion that is initially confined to a
particular orbital plane can be unstable to motion in the
perpendicular direction [AB].  The equation that describes the
development of this instability takes the form of equation
(\ref{eq:basic}).  Further, the motion in the original orbital plane
often displays chaotic behavior, which becomes more extreme as the
axis ratios of the potential increase [BT].  In this application, the
motion in the original orbit plane -- in particular, the distance to
the center of the coordinate system -- determines the magnitude of the
forcing strength $q_k$ that appears in Hill's equation.  The crossing
time, which varies from orbit to orbit, determines the value of the
oscillation parameter $\af_k$. As a result, the chaotic behavior in
the original orbital plane leads to random forcing effects in the
differential equation that determines instability of motion out of the
plane (see Appendix A for further discussion). 

Given that Hill's equations arise in a wide variety of physical
problems [MW], we expect that applications with random forcing terms
will be common. Although the literature on
stochastic differential equations 
is vast (e.g., see the review of [BL]), specific results
regarding Hill's equations with random forcing terms are relatively
rare.

In this application, Hill's equation is periodic or nearly periodic
(we generalize to the case of varying periods for the basic cycles),
and the forcing strength $q_k$ varies from cycle to cycle. Since the
forcing strength is fixed over a given cycle, one can solve the Hill's
equation for each cycle using previously developed methods [MW], and
then match the solutions from cycle to cycle using a discrete map. As
shown below, the long-time solution can be developed by repeated
multiplication of 2 $\times$ 2 matrices that contain a random
component in their matrix elements. 

The subject of random matrices, including the long term behavior of
their products, is also the subject of a great deal of previous work
[BL, DE, BD, FK, FU, LR, ME, VI].  In this application, however,
Hill's equation determines the form of the random matrices and the
repeated multiplication of this type of matrix represents a new and
specific application.  Given that instances where analytic results can
be obtained are relatively rare, this set of solutions adds new
examples to the list of known cases.

This paper is organized as follows. In \S 2, we present the basic
formulation of the problem, define relevant quantities, and show that
aperiodic generalizations of the problem can be reduced to random
Hill's equations. The following section (\S 3) presents the main
results of the paper: We find specific results regarding the growth
rates of instability for random Hill's equations in the limit of large
forcing strengths (i.e., in the limit where the equations are robustly
unstable). These results are presented for purely positive and for
mixed signs in the $2 \times 2$ matrix map. We also find limiting
forms and constraints on the growth rates.  Finally, we find bounds
and estimates for the errors incurred by working in the limit of large
forcing strengths.  This work is related to the general existence
results of [FU] but provides much more detailed information in our
setting.  In the next section (\S 4) we consider the limit where the
forcing terms are Dirac delta functions; this case allows for analytic
solutions to the original differential equation. We note that the
growth rates calculated here (\S 3) depend on the distribution of the
ratios of the principal solutions to equation (\ref{eq:basic}), rather
than (directly) on the distributions of the parameters $(\af_k, q_k)$. 
Using the analytic solutions for the delta function limit (\S 4), we
thus gain insight into the transformation between the distributions of
the input parameters $(\af_k, q_k)$ and the parameters that specify
the growth rates.  Finally, we conclude, in \S 5, with a summary and
discussion of our results.

\section{FORMULATION}  

\noindent
{\bf Definition:} A {\it Random Hill's equation} is defined here to be
of the form given by equation (\ref{eq:basic}) where the forcing
strength $q_k$ and oscillation parameter $\af_k$ vary from cycle to
cycle. Specifically, the parameters $q_k$ and $\af_k$ are stochastic
variables that take on new values every cycle $0 \le \tmod \le \pi$,
and the values are sampled from known probability distributions $P_q
(q)$ and $P_\af (\af)$.

\subsection{Hill's Equation with Fixed Parameters} 

Over a single given cycle, a random Hill's equation is equivalent to
an ordinary Hill's equation and can be solved using known methods [MW].

\noindent 
{\bf Definition:} The {\it growth factor} $f_c$ per cycle (the Floquet
multiplier) is given by the solution to the characteristic equation
and can be written as
\be 
f_c = { \discrim + (\discrim^2 - 4)^{1/2} \over 2} \, , 
\ee
where the discriminant $\discrim = \discrim (q, \af)$ is defined by 
\be 
\discrim \equiv y_1 (\pi) + {dy_2 \over dt}(\pi) \, ,
\label{eq:discrim} 
\ee 
and where $y_1$ and $y_2$ are the principal solutions [MW]. 

It follows from Floquet's Theorem that $|\discrim| > 2$ is a
sufficient condition for instability [MW, AS]. In addition, 
periodic solutions exist when $|\discrim|$ = 2. 

\subsection{Random Variations in Forcing Strength} 

We now generalize to the case where the forcing strength $q_k$ and
oscillation parameter $\af_k$ vary from cycle to cycle. In other
words, we consider each period from $t=0$ to $t=\pi$ as a cycle, and
consider the effects of successive cycles with varying values of
$(q_k, \af_k)$.

During any given cycle, the solution can be written as a linear
combination of the two principal solutions $y_1$ and $y_2$.  Consider
two successive cycles. The first cycle has parameters $(q_a, \af_a)$
and solution
\be 
f_a (t) = \alpha_a y_{1a} (t) + \beta_a y_{2a} (t) \, , 
\ee 
where the solutions $y_{1a}(t)$ and $y_{2a}(t)$ correspond to those
for an ordinary Hill's equation when evaluated using the values 
$(q_a, \af_a)$.  Similarly, for the second cycle with parameters
$(q_b, \af_b)$ the solution has the form 
\be 
f_b (t) = \alpha_b y_{1b} (t) + \beta_b y_{2b} (t) \, .  
\ee 
Next we note that the new coefficients $\alpha_b$ and $\beta_b$ are 
related to those of the previous cycle through the relations 
\be
\alpha_b = \alpha_a y_{1a} (\pi) + \beta_a y_{2a} (\pi) 
\qquad {\rm and} \qquad \beta_b = 
\alpha_a {d y_{1a} \over dt} (\pi) + 
\beta_a {d y_{2a} \over dt} (\pi) \, . 
\label{eq:alphabeta} 
\ee
The new coefficients can thus be considered as a two dimensional
vector, and the transformation between the coefficients in one cycle
and the next cycle is a $2 \times 2$ matrix. Here we consider the case
in which the equation is symmetric with respect to the midpoint $t =
\pi/2$. This case arises in the original orbit problem that motivated
this study --- the forcing function is determined by the orbit as it
passes near the center of the potential and this passage is symmetric
(or very nearly so). It also makes sense to consider the symmetric 
case, which is easier, first. Since the Wronskian of the original 
differential equation is unity, the number of independent matrix
coefficients is reduced further, from four to two. We thus have 
the following result: 

\noindent 
{\bf Proposition 1:} {\it The transformation between the coefficients
$\alpha_a, \beta_a$ of one cycle and the coefficients $\alpha_b,
\beta_b$ of the next may be written in the form} 
\be 
\left[ \matrix{ \alpha_b \cr \beta_b \cr } \right] = 
\left[ \matrix{ h & (h^2 -1)/g \cr g & h \cr } \right] 
\left[ \matrix{ \alpha_a \cr \beta_a \cr } \right] \equiv {\bf M} (q_a) 
\left[ \matrix{ \alpha_a \cr \beta_a \cr } \right] \, , 
\label{eq:matrixdef} 
\ee
{\it where the matrix $\bf M$ (defined in the second equality) depends on
the values $(q_a, \af_a)$ and $h = y_1 (\pi)$ and $g = {\dot
y}_1 (\pi)$ for a given cycle. } 

\noindent
{\bf Proof:} This result can be  verified by standard matrix multiplication, 
which yields equation (\ref{eq:alphabeta}) above. $\thmbox$ 

After $N$ cycles with varying values
of $(q_k, \af_k)$, the solution retains the general form given above, 
where the coefficients are determined by the product of matrices 
according to
\be 
\left[ \matrix{ \alpha_N \cr \beta_N \cr } \right] = {\bf M}^{(N)} 
\left[ \matrix{ \alpha_0 \cr \beta_0 \cr } \right] 
\qquad {\rm where} \qquad 
{\bf M}^{(N)} \equiv \prod_{k=1}^N {\bf M}_k (q_k, \af_k) \, . 
\label{eq:product} 
\ee

This formulation thus transforms the original differential equation
(with a random element) into a discrete map. The properties of the
product matrix ${\bf M}^{(N)}$ determine whether the solution is
unstable and the corresponding growth rate.

\noindent 
{\bf Definition:} The {\it asymptotic growth rate} $\gamma_\infty$ is
that experienced by the system when each cycle amplifies the growing
solution by the growth factor appropriate for the given value of the
forcing strength for that cycle, i.e.,
\be 
\gamma_\infty \equiv \lim_{N \to \infty} 
{1 \over \pi N} \log \left[ \prod_{k=1}^N {1 \over 2} 
\bigl\{ \discrim_k + \sqrt{\discrim_k^2 - 4} \bigr\} \right]  \, ,  
\label{eq:gaminf} 
\ee 
where $\discrim_k = \discrim (q_k, \af_k)$ is defined by equation
(\ref{eq:discrim}), and where this expression is evaluated in the
limit $N \to \infty$.  In this definition, it is understood that if
$|\discrim_k| < 2$ for a particular cycle, then the growth factor is
unity for that cycle, resulting in no net contribution to the product
(for that cycle).

Notice that the factor of $\pi$ appears in this definition of the
growth rate because the original Hill's equation is taken to be
$\pi$-periodic [MW, AS]. As we show below, the growth rates of the
differential equation are determined by the growth rates resulting
from matrix multiplication. In many cases, however, the growth rates
for matrix multiplication are given without the factor of $\pi$ [BL,
FK], so there is a mismatch of convention (by a factor of $\pi$)
between growth rates of Hill's equations and growth rates of matrix
multiplication.

Notice that this expression for the asymptotic growth rate 
takes the form 
\be 
\gamma_{\infty} = \lim_{N \to \infty} {1 \over N} 
\sum_{k=1}^N \gamma (q_k, \af_k) \to  \langle \gamma \rangle \, , 
\label{eq:gamtwo} 
\ee
where $\gamma(q_k, \af_k)$ is the growth rate for a given cycle. The
asymptotic growth rate is thus given by the expectation value of the
growth rate per cycle for a given probability distribution for the
parameters $q_k$ and $\af_k$.

We note that a given system does not necessarily experience growth at
the rate $\gamma_\infty$ because the solutions must remain continuous
across the boundaries between subsequent cycles.  This requirement
implies that the solutions during every cycle will contain an
admixture of both the growing solution and the decaying solution for
that cycle, thereby leading to the possibility of slower growth. In
some cases, however, the growth rate is larger than $\gamma_\infty$,
i.e., the stochastic component of the problem aids and abets the
instability. One could also call $\gamma_\infty$ the direct growth rate. 

\subsection{Generalization to Aperiodic Variations} 

\noindent 
{\bf Theorem 1:} {\it Consider a generalization of Hill's equation so that
the cycles are no longer exactly $\pi$-periodic. Instead, each cycle
has period $\mu_k \pi$, where $\mu_k$ is a random variable that
averages to unity. Then variations in period are equivalent to
variations in $(q,\af)$, i.e., the problem with three stochastic variables
$(q_k, \af_k, \mu_k)$ reduces to a $\pi$-periodic problem with only two
stochastic variables $(q_k, \af_k)$.} 

\noindent 
{\bf Proof:} With this generalization, the equation of motion 
takes the form
\be
{d^2 y \over dt^2} + 
\left[ \af_k + q_k {\hat Q} (\mu_k t) \right] y = 0 \, , 
\ee 
where we have normalized the forcing frequency to have unit amplitude
($\hat Q$ = $Q/q_k$). Since $\hat Q$ (and $Q$) are $\pi$-periodic, the
$jth$ cycle is defined over the time interval $0 \le \mu_k t \le \pi$,
or $0 \le t \le \pi/\mu_k$. We can re-scale both the time variable and
the ``constants'' according to
\be 
t \to \mu_k t \, , \qquad 
\af_k \to \af_k / \mu_k^2 = \lamwig \, , \qquad 
{\rm and} \qquad q_k \to q_k/\mu_k^2 = \qwig \, ,  
\ee 
so the equation of motion reduces to the familiar form 
\be 
{d^2 y \over dt^2} + 
\left[ \lamwig + \qwig {\hat Q} (t) \right] y = 0 \, . 
\ee 
Thus, the effects of varying period can be incorporated into 
variations in the forcing strength $q_k$ and oscillation 
parameter $\af_k$. $\thmbox$ 

\section{HILL'S EQUATION IN THE UNSTABLE LIMIT} 

In this section we consider Hill's equation in general form (for the
delta function limit see \S 4), but restrict our analysis to the case
of symmetric potentials so that $y_1 (\pi) = h = {\dot y}_2 (\pi)$. We
also consider the {\it highly unstable limit}, where we define this
limit to correspond to large $h \gg 1$.  Since the 2 $\times$ 2 matrix
of the discrete map must have its determinant equal to unity, the
matrix of the map has the form given by equation (\ref{eq:matrixdef}),
where the values of $h$ and $g$ depend on the form of the forcing
potential.  

The discrete map can be rewritten in the general form 
\be
{\bf M} = h \left[ \matrix{1 & x \cr 1/x & 1} \right] \, + 
\left[ \matrix{0 & -1/g \cr 0 & 0} \right] \,  \, . 
\label{eq:intermediate} 
\ee
In the highly unstable limit $h \to \infty$, the matrix 
simplifies to the approximate form
\be
{\bf M} \approx h \left[ \matrix{1 & x \cr 1/x & 1} \right] \, 
\equiv h {\bf C} \, , 
\label{eq:genmap}
\ee
where we have defined $x \equiv h/g$, and where the second equality
defines the matrix $\bf C$.  

In this problem we are concerned with both the long-time limit $N \to
\infty$, and the ``unstable'' limit $h \to \infty$.  In the first
instance considered here we take the unstable limit first, but below
we analyze precisely the difference between taking the long time limit
first and then the unstable limit.

\subsection{Fixed Matrix of the Discrete Map} 

The simplest case occurs when the stochastic component can be
separated from the matrix, i.e., when the matrix ${\bf C}$ does not vary
from cycle to cycle. This case arises when the Hill's equation does
not contain a random component; it also arises when the random
component can be factored out so that $x$ does not vary from cycle to
cycle, although the leading factors $h_k$ can vary.  In either case,
the matrix $\bf C$ is fixed.  Repeated multiplications of the matrix
$\bf C$ are then given by
\be
{\bf C}^N = 2^{N-1} {\bf C} \, . 
\ee
With this result, after $N$ cycles the Floquet multiplier 
(eigenvalue) of the product matrix and the corresponding growth 
rate take the form 
\be
\Lambda = \prod_{k=1}^N (2 h_k) 
\qquad {\rm and} \qquad \gamma = \lim_{N \to \infty} 
{1 \over \pi N} \sum_{k=1}^N \log(2h_k) \, . 
\label{eq:const}
\ee
Note that this result applies to the particular case of Hill's
equation in the delta function limit (\S 4), where the forcing
strength $q_k$ varies from cycle to cycle but the frequency parameter
$\af_k$ is constant.  The growth rate in equation (\ref{eq:const})
is equal to the asymptotic growth rate $\gamma_\infty$
(eq. [\ref{eq:gaminf}]) for this case.

\subsection{General Results in the Unstable Limit} 

We now generalize to the case where the parameters of the differential
equation vary from cycle to cycle.  For a given cycle, the discrete
map is specified by a matrix of the form specified by equation
(\ref{eq:genmap}), where $x = x_k = h_k/g_k$, with varying values from
cycle to cycle. The values of $x_k$ depend on the parameters
$(q_k, \af_k)$ through the original differential equation.  After $N$
cycles, the product matrix ${\bf M}^{(N)}$ takes the form
\be
{\bf M}^{(N)} = \prod_{k=1}^N h_k \, \, 
\prod_{k=1}^N {\bf C}_k \, \, ,
\ee 
where we have separated out the two parts of the problem. 
One can show (by induction) that the product of $N$ 
matrices ${\bf C}_k$ have the form 
\be 
{\bf C}^{(N)} = \prod_{k=1}^N {\bf C}_k = 
\left[ \matrix{ \sumtn & x_1 \sumtn \cr
\sumbn/x_1 & \sumbn } 
\right] \, , 
\label{eq:productck} 
\ee
where $x_1$ is the value of the variable for the 
first cycle and where the sums $\sumtn$ and $\sumbn$ 
are given by 
\be
\sumtn = \sum_{j=1}^{2^{N-1}} r_j 
\qquad {\rm and} \qquad 
\sumbn = \sum_{j=1}^{2^{N-1}} {1 \over r_j} \, ,  
\ee
where the variables $r_j$ are ratios of the form 
\be 
r_j = {x_{a_1} x_{a_2} \dots x_{a_n} \over 
x_{b_1} x_{b_2} \dots x_{b_n} } \, . 
\label{eq:ratdef} 
\ee
The ratios $r_j$ arise from repeated multiplication of the matrices
${\bf C}_k$, and hence the indices lie in the range $1 \le a_i, b_i
\le N$.  The $r_j$ always have the same number of factors in the
numerator and the denominator, but the number of factors ($n$) varies
from 0 (where $r_j = 1$) up to $N/2$. This upper limit arises because 
each composite ratio $r_j$ has $2 n$ values of $x_j$, which must all 
be different, and because the total number of possible values is $N$. 

Next we define a composite variable 
\be
\swig \equiv {1 \over 2^N} \bigl[ \sumtn + \sumbn \bigr]   
= {1 \over 2^N} \sum_{j=1}^{2^{N-1}} \bigl( r_j + 
{1 \over r_j} \bigr) \, . 
\label{eq:swigdef} 
\ee 
With this definition, the (growing) eigenvalue $\Lambda$ of the product 
matrix ${\bf M}^{(N)}$ takes the form 
\be 
\Lambda = \swig \, \prod_{k=1}^N (2 h_k) 
\ee
and the corresponding growth rate of the instability 
has the form 
\be 
\gamma = \lim_{N \to \infty} \left[ 
{1 \over \pi N} \sum_{k=1}^N \log(2 h_k) 
+  {1 \over N \pi} \log \swig \right] \, .  
\ee
The first term is the asymptotic growth rate $\gamma_\infty$ defined
by equation (\ref{eq:gaminf}) and is thus an average of the growth
rates for the individual cycles.  All of the additional information
regarding the stochastic nature of the differential equation is
encapsulated in the second term through the variable $\swig$. For
example, if the composite variable $\swig$ is finite in the limit $N
\to \infty$, then the second term would vanish. As shown below,
however, the stochastic component can provide a significant
contribution to the growth rate, and can provide either a stabilizing
or destabilizing influence. In the limit $N \to \infty$, we can thus
write the growth rate in the from
\be 
\gamma = \gamma_\infty + \Delta \gamma \, , 
\label{eq:gamsum} 
\ee
where we have defined the correction term $\Delta \gamma$, 
\be
\Delta \gamma \equiv \lim_{N \to \infty} 
{1 \over N \pi} \log \swig \, . 
\ee 

Since the asymptotic growth rate $\gamma_\infty$ is straightforward to
evaluate, the remainder of this section focuses on evaluating $\Delta
\gamma$, as well as finding corresponding estimates and constraints.
This correction term $\Delta \gamma$ is determined by the discrete map
$\bf C$, whose matrix elements are given by the ratios $x = h/g$,
where $h$ and $g$ are determined by the solutions to Hill's equation
over one cycle. One should keep in mind that the parameters in the
original differential equation are $(\af_k, q_k)$. The distribution of
these parameters determines the distributions of the principal
solutions (the distributions of $h_k$ and $g_k$), whereas the
distribution of the ratios $x_k$ of these latter quantities determines
the correction $\Delta \gamma$ to the growth rate. The problem thus
separates into two parts: [1] The transformation between the
distributions of the parameters $(\af_k, q_k)$ and the resulting
distribution of the ratios $x_k$ that define the discrete map, and [2]
The growth rate of the discrete map for a given distribution of $x_k$.
The following analysis focuses on the latter issue (whereas \S 4
provides an example of the former issue).

\subsection{Growth Rates for Positive Matrix Elements} 

This subsection addresses the cases where the ratios $x_k$ that define
the discrete map ${\bf C}$ all have the same sign. For this case, the
analysis is simplified, and a number of useful results can be
obtained.

\bigskip 
\noindent 
{\bf Theorem 2:} {\it Consider the general form of Hill's equation in the
unstable limit so that $h = y_1 (\pi) = {\dot y}_2 (\pi) \gg 1$.  For
the case of positive matrix elements, $r_j > 0$, the growth rate is
given by equation (\ref{eq:gamsum}) where the correction term 
$\Delta \gamma$ is given by} 
\be 
\Delta \gamma = \lim_{N \to \infty} {1 \over \pi N}  
\sum_{j=1}^N \log (1 + x_{j1}/x_{j2} ) \, - 
{\log 2 \over \pi} \, , 
\label{eq:thm2} 
\ee
{\it where $x_{j1}$ and $x_{j2}$ represent two different (independent)
samples of the $x_j$ variable.}\footnote{Specifically, the index $j$
labels the cycle number, and the indices $j1$ and $j2$ label two
successive samples of the $x$ variable; since the stochastic
parameters of the differential equations are assumed to be independent
from cycle to cycle, however, the variables $x_{j1}$ and $x_{j2}$ can
be any independent samples.}

\noindent 
{\bf Proof:} Using the same induction argument that led to equation
(\ref{eq:productck}), one finds that from one cycle to the next the
sums $\sumtn$ and $\sumbn$ vary according to
\be 
\Sigma_{T(N+1)} = \sumtn + {x \over x_0} \sumbn \, , 
\ee
and
\be 
\Sigma_{B(N+1)} = \sumbn + {x_0 \over x} \sumtn \, . 
\ee
In this notation, the variable $x$ (no subscript) represents the value
of the $x$ variable at the current cycle, whereas $x_0$ represents the
value at the initial cycle.  The growing eigenvalue of the product
matrix of equation (\ref{eq:productck}) is given by $\Lambda = \sumtn
+ \sumbn$. As a result, the eigenvalue (growth factor) varies from
cycle to cycle according to 
\be
\Lambda^{(N+1)} = \Lambda^{(N)} + 
{x \over x_0} \sumbn + {x_0 \over x} \sumtn \, = 
\Lambda^{(N)} \left[ 1 + { (x / x_0) \sumbn + (x_0 / x) \sumtn 
\over \sumbn + \sumtn } \right] \, . 
\ee 
The overall growth factor is then determined by the product 
\be 
\Lambda^{(N)} = \prod_{j=1}^N \left[ 1 + 
{ (x/x_0) \sumbn + (x_0/x) \sumtn \over 
\sumbn + \sumtn} \right]  \, .
\ee
The growth rate of matrix multiplication is determined by 
setting the above product equal to $\exp[N \pi \gamma]$. 
The growth rate $\Delta \gamma$ also includes the factor 
of 2 per cycle that is included in the definition of the 
asymptotic growth rate $\gamma_\infty$. We thus find that 
\be
\Delta \gamma \approx {1 \over N \pi} \sum_{j=1}^N \log 
\left[ 1 + {(x_{j1}/x_{j2}) \sumbn + (x_{j2}/x_{j1}) 
\sumtn \over \sumbn + \sumtn } \right] \, - 
{\log 2 \over \pi} \, . 
\label{eq:almost} 
\ee 
Note that this expression provides the correction $\Delta \gamma$ to
the growth rate. The full growth rate is given by $\gamma$ =
$\gamma_\infty$ + $\Delta \gamma$ (where $\gamma_\infty$ is specified
by eq. [\ref{eq:gaminf}] and $\Delta \gamma$ is specified by
eq. [\ref{eq:thm2}]). In the limit of large $N$, the ratio of the 
sums $\sumtn$ and $\sumbn$ approaches unity, almost surely, so that 
\be
{\sumtn \over \sumbn} \to 1 \qquad {\rm as} \qquad N \to \infty \, . 
\label{eq:ratio} 
\ee 
This result follows from the definition of $\sumtn$ and $\sumbn$: The
terms in each of these two sums is the product of ratios $x_a/x_b$,
and, the terms $r_j$ in the first sum $\sumtn$ are the inverse of
those ($1/r_j$) in the second sum $\sumbn$. Since the fundamental
variables $x_k$ that make up these ratios, and the products of these
ratios, are drawn from the same distribution, the above condition
(\ref{eq:ratio}) must hold.  As a consequence, the expression for the
growth rate given by equation (\ref{eq:almost}) approaches that of
equation (\ref{eq:thm2}).  $\thmbox$

\noindent
{\bf Corollary 2.1:} {\it Let $\sigma_0$ be the variance of the
composite variable $\log(x_{j1}/x_{j2})$ (see Theorem 2). The
correction to the growth rate is positive semi-definite; specifically,
$\Delta \gamma \ge 0$ and $\Delta \gamma \to 0$ in the limit $\sigma_0
\to 0$. Further, in the limit of small variance, the growth rate
approaches the asymptotic form $\Delta \gamma \to \sigma_0^2 / (8\pi)$.} 

\noindent 
{\bf Proof:} 
In the limit of small $\sigma_0$, we can write $x_j = 1 + \delta_j$,
where $|\delta_j| \ll 1$. In this limit, equation (\ref{eq:thm2}) for
the growth rate becomes
\be 
\Delta \gamma = \lim_{N \to \infty} {1 \over \pi N} \sum_{j=1}^N 
\log \bigl[ 2 + \delta_{j1} - \delta_{j2} + \delta_{j2}^2 - 
\delta_{j1} \delta_{j2} + {\cal O} (\delta^3) \bigr] \, 
- {\log 2 \over \pi} \, . 
\ee 
In the limit $|\delta_j| \ll 1$, we can expand the logarithm, 
and the above expression simplifies to the form 
\be 
\Delta \gamma = \lim_{N \to \infty} {1 \over 2 \pi N} \sum_{j=1}^N 
\bigl[ \delta_{j1} - \delta_{j2} + \delta_{j2}^2 - 
\delta_{j1} \delta_{j2} - (\delta_{j1} - \delta_{j2})^2 / 4  
+ {\cal O} (\delta^3) \bigr] \, . 
\ee 
Evaluation of the above expression shows that 
\be
\Delta \gamma = {1 \over 2 \pi} 
\left[ \langle \delta_{j2}^2 \rangle - {1 \over 4} 
\langle (\delta_{j1} - \delta_{j2})^2 \rangle 
+ {\cal O} (\delta^3) \right] \, \to {\sigma_0^2 \over 8 \pi} . 
\ee 
As a result, $\Delta \gamma \ge 0$. In the limit $\sigma_0 \to 0$, 
all of the $x_j$ approach unity and $\delta_j \to 0$; therefore, 
$\Delta \gamma \to 0$ as $\sigma_0 \to 0$. $\thmbox$ 

\medskip 

Although equation (\ref{eq:thm2}) is exact, the computation 
of the expectation value can be difficult in practice. As a 
result, it is useful to have simple constraints on the growth 
rate in terms of the variance of the probability distribution 
for the variables $x_k$. In particular, a simple bound can be 
derived: 

\noindent
{\bf Theorem 3:} {\it Consider the general form of Hill's equation in
the unstable limit so that $h = y_1 (\pi) = {\dot y}_2 (\pi) \gg
1$. Take the variables $r_j > 0$.  Then the growth rate is given by
equation (\ref{eq:gamsum}) and the correction term $\Delta \gamma$
obeys the constraint}
\be 
\Delta \gamma \le {\sigma_0^2 \over 4 \pi} \, , 
\label{eq:dgbound} 
\ee 
{\it where $\sigma_0^2$ is the variance of the distribution of the
variable $\xi = \log (x_{j1}/x_{j2})$, and where $x_j$ are independent
samplings of the ratios $x_j = h_j/g_j$.} 

\noindent 
{\bf Proof:} First we define the variable $\xi_j = \log r_j$, where $r_j$ 
is given by equation (\ref{eq:ratdef}) above with a fixed value of $n$.
In the limit of large $n$, the variable $\xi_j$ has zero mean and will
be normally distributed. If the variables $x_j$ are independent, the
variance of the composite variable $\xi_j$ will be given by
\be
\sigma_\xi^2 = n \sigma_0^2 \, . 
\label{eq:varcom} 
\ee 
As shown below, is order to obtain $2^N$ terms in the sums $\sumtn$ 
and $\sumbn$, almost all of the variables $r_j$ will fall in the 
large $n$ limit; in addition, $n \to \infty$ in the limit $N \to
\infty$.  As a result, we can consider the large $n$ limit to be valid
for purposes of evaluating the correction term $\Delta \gamma$. In
practice, the variables will not be completely independent, so the
actual variance will be smaller than that given by equation
(\ref{eq:varcom}); nonetheless, this form can be used to find the 
desired upper limit.

Given the large $n$ limit and a log-normal distribution of $r_j$, the
expectation values $\langle r_j \rangle$ and $\langle 1/r_j \rangle$ 
are given by
\be
\langle r_j \rangle = \exp\bigl[ n \sigma_0^2 / 2 \bigr] = 
\langle 1 / r_j \rangle \, . 
\label{eq:rjexpect} 
\ee
Note that the variable $\xi_j$ is normally distributed, and we are
taking the expectation value of $r_j = \exp \xi_j$; since the mean of
the exponential is not necessarily equal to the exponential of the
mean, the above expression contains the (perhaps counterintuitive)
factor of 2.  As expected, larger values of $n$ allow for a wider
possible distribution and result in larger expectation values. The
maximum expectation values thus occur for the largest values of
$n$. Since $n < N/2$, these results, in conjunction with equation
(\ref{eq:swigdef}) imply that $\swig$ obeys the constraint 
\be 
\swig < \exp \bigl[ N \sigma_0^2 / 4 \bigr] \, . 
\ee 
The constraint claimed in equation (\ref{eq:dgbound}) 
then follows immediately. 

{\sl Combinatorics:} To complete the argument, we must show that most
of the variables $r_j$ have a large number $n$ of factors (in the
limit of large $N$).  The number of terms in the sums $\sumtn$ and
$\sumbn$ is large, namely $2^{N-1}$. Further, the ratios $r_j$ must
contain $2n$ different values of the variables $x_k$. The number
$P(n|N)$ of different ways to choose the $2n$ variables for $N$ cycles
(and hence $N$ possible values of $x_k$) is given by the expression
\be 
P(n|N) = {N! \over (N - 2n)! (n!)^2 } \, . 
\ee
Notice that this expression differs from the more familiar binomial
coefficient because the values of $r_j$ depend on whether or not the
$x_k$ factors are in the numerator or denominator of the ratio $r_j$.
Next we note that if $n \ll N$, then the following chain of inequalities 
holds for large $N$: 
\be 
P(n|N) < {N^{2n} \over (n!)^2 } \ll 2^{N-1} \, . 
\ee 
For large $N$ and $n \ll N$, the central expression increases like a
power of $N$, whereas the right hand expression increases
exponentially with $N$. As a result, for $n \ll N$, there are not
enough different ways to choose the $x_k$ values to make the required
number of composite ratios $r_j$. In order to allow for enough
different $r_j$, the number $n$ of factors must be large (namely,
large enough so that $n \ll N$ does not hold) for most of the $r_j$.
This conclusion thus justifies our use of the large $n$ limit in the
proof of Theorem 3 (where we used a log-normal form for the composite
distribution to evaluate the expectation values $\langle r_j \rangle$
and $\langle 1/r_j \rangle$). $\thmbox$ 

{\sl Estimate:} Theorem 3 provides an upper bound on the contribution
of the correction term $\Delta \gamma$ to the overall growth rate.
This bound depends on the value of $n$, which determines the magnitude
of the expectation value $\langle r_j \rangle$. It is useful to have
an estimate of the ``typical'' size of $n$.  In rough terms, the value
of $n$ must be large enough so that the number of possible
combinations is large enough to account for the $2^{N-1}$ terms in the
sums $\sumtn$ and $\sumbn$. For each $n$, we have $P(n|N)$
combinations. As a rough approximation, $n P(n|N)$ accounts for all of
the combinations of size less than $n$.  If we set $n P(n|N) = 2^N$,
we can solve for the ratio $n/N$ required to have enough terms, and
find $n/N \approx 0.11354 \dots \approx 1/9$.  As a result, we expect
the ratio $n/N$ to lie in the range
\be
{1 \over 9} < {n \over N} < {1 \over 2} \, . 
\ee 
If we use this range of $n/N$ to evaluate the expectation value using
equation (\ref{eq:rjexpect}), and estimate the growth rate, the upper
end of this range provides a rigorous upper bound (Theorem 3). The
lower end of the range only represents a rough guideline, however,
since the variables are not fully independent. Nonetheless, it can be
used to estimate the expectation values $\langle r_j \rangle$.

\begin{figure}
\figurenum{1}
{\centerline{\epsscale{0.90} \plotone{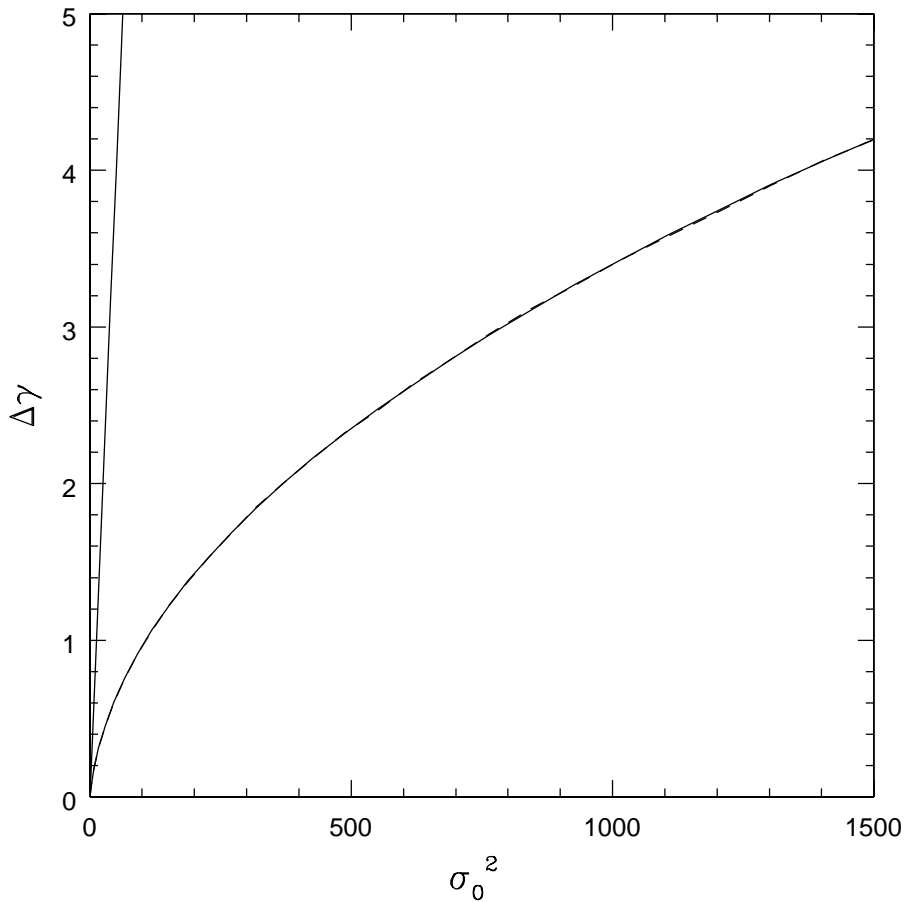} }} 
\figcaption{ Comparison of the bound of Theorem 3 and the prediction
of Theorem 2 with results from numerical experiments.  All cases use
matrices ${\bf C}_k$ of the form given by equation (\ref{eq:genmap}),
where the variables $x_k$ are chosen according to distributions with
variance $\sigma_0^2$.  For each distribution, the growth rate $\Delta
\gamma$ due to matrix multiplication is plotted versus the variance of
the distribution of the composite variable $\xi = \log(x_j/x_k)$,
where $x_k = y_{1k} (\pi) / {\dot y}_{1k} (\pi)$ and, similarly,  
$x_j = y_{1j} (\pi) / {\dot y}_{1j} (\pi)$. The solid curve 
shows the results obtained by averaging together 1000 realizations of
the numerical experiments; the overlying dashed curve shows the
prediction of Theorem 2. The straight solid line shows the upper bound
of Theorem 3, i.e., $\Delta \gamma \le \sigma_0^2/(4 \pi)$. }
\label{fig:gamlimit} 
\end{figure}

Notice that the upper bound is conservative. Figure \ref{fig:gamlimit}
shows a comparison of the actual growth rate (from Theorem 2) and the
bound (Theorem 3). At large variance, the actual growth rate is much 
less than our bound. In fact, as shown in the following section, in 
the limit of large variance, the growth rate $\Delta \gamma \propto 
\sigma_0$ (rather than $\propto \sigma_0^2$). 

For this numerical experiment, we used a particular form for the $x_k$
variables, namely $x_k = 0.01 + (10 a \xi_k)^a$, where $\xi_k$ is a
random variable in the range $0 \le \xi_k \le 1$ and $a$ is a
parameter that is chosen to attain varying values of $\sigma_0^2$. The
exact form of the curve $\Delta \gamma (\sigma_0^2)$ depends on the
distribution of the $x_k$. However, all of the distributions studied
result in the general form shown in Figure \ref{fig:gamlimit}, and all
of the cases show the same agreement between numerical experiments and
the predictions of Theorem 2.

\subsection{Error Bounds and Estimates} 

The analysis presented thus far is valid in the highly unstable limit,
as defined at the beginning of this section.  In other words, we have
found an exact solution to the reduced problem, as encapsulated in
equation (\ref{eq:genmap}). In this problem we are taking two limits,
the long-time limit $N \to \infty$ and the ``unstable'' limit $h \to
\infty$. In the reduced problem, as analyzed above, we take the limit
$h \to \infty$ first, and then consider the long-time limit $N \to
\infty$. In this subsection, we consider the accuracy of this approach
by finding bounds (and estimates) for the errors in the growth rates
incurred from working in the highly unstable limit. In other words, we
find bounds on the difference between the results for the full problem
(with large but finite $h_k$) and the reduced problem. 

To assess the error budget, we write the general matrix (for the 
full problem) in the form  
\be
{\bf M} = h {\bf B} \qquad {\rm where} \qquad 
{\bf B} \equiv \left[ \matrix{1 & x \fsub \cr 1/x & 1} \right] \, . 
\label{eq:full} 
\ee
This form is the same as the matrix of the reduced problem (in the
unstable limit) except for the correction factor $\fsub$ in the (1,2)
matrix element, where $\fsub \equiv (1 - 1/h^2)$. 

Let $(\Delta \gamma)_B$ denote the growth rate for the matrix $\bf B$
for the full problem defined in equation (\ref{eq:full}).  Similarly,
let $(\Delta \gamma)_C$ denote the growth rate found previously for
the reduced problem using the matrix $\bf C$ defined in equation
(\ref{eq:genmap}).  Through repeated matrix multiplications, the
product of matrices ${\bf B}_k$ will be almost the same as for the
product of matrices ${\bf C}_k$, where the difference is due to the
continued accumulation of factors $\fsubk$. Note that the index $k$,
as introduced here, denotes the cycle number, and that all of these
quantities vary from cycle to cycle.

\noindent
{\bf Proposition 2:} {\it The error $\varepsilon_{BC} = (\Delta \gamma)_C -
(\Delta \gamma)_B$ introduced by using the reduced form of the problem
(the matrices ${\bf C}_k$) instead of the full problem (the matrices
${\bf B}_k$) is bounded by}
\be
0 < \varepsilon_{BC} < - {1 \over \pi} \langle \log \fsubk \rangle \, . 
\ee 

\noindent 
{\bf Proof:} Since $\fsubk < 1$, by definition, we see immediately
that the growth rate for the full problem is bounded from above by
that of the reduced problem, i.e.,
\be
(\Delta \gamma)_B < (\Delta \gamma)_C \, . 
\label{no1}
\ee 
Next we construct a new matrix of the form 
\be
{\bf A} \equiv \fsub \left[ \matrix{1 & x \cr 1/x & 1} \right] \, 
= \fsub {\bf C} \, . 
\label{eq:bound} 
\ee
The products of the matrices ${\bf A}_k$ will be almost the same as
those for the matrices ${\bf B}_k$, where the difference is again due
to the inclusion of additional factors of $\fsubk$. Since the $\fsubk
< 1$, we find that the growth rate for this benchmark problem is less
than (or equal to) that of the full problem, i.e., $(\Delta \gamma)_A
< (\Delta \gamma)_B$.  Further, the growth rate $(\Delta \gamma)_A$
for this new matrix can be found explicitly and is given by
\be
(\Delta \gamma)_A = (\Delta \gamma)_C + \lim_{N \to \infty} 
{1 \over \pi N} \log \left[ \prod_{k=1}^N \fsubk \right] = 
(\Delta \gamma)_C + \lim_{N \to \infty} 
{1 \over \pi N} \sum_{k=1}^N \log \fsubk \, .  
\label{no2}
\ee
Combining equations (\ref{no1})  and (\ref{no2})
shows that the growth rate for 
the full problem $(\Delta \gamma)_B$ is bounded on both sides and 
obeys the constraint  
\be
(\Delta \gamma)_C + {1 \over \pi} 
\langle \log \fsubk \rangle < (\Delta \gamma)_B < (\Delta \gamma)_C \, . 
\ee 
Notice that the expectation value $\langle \log \fsubk \rangle < 0$
since $\fsubk < 1$. The error $\varepsilon_{BC}$ introduced by using
the reduced form of the problem (the matrices ${\bf C}_k$) instead of
the full problem (the matrices ${\bf B}_k$) is thus bounded by 
\be
0 < \varepsilon_{BC} < - {1 \over \pi} \langle \log \fsubk \rangle \, . 
\ee 

This bound can be made tighter by a factor of 2. Note that the product
of two matrices of the full problem has the form
\be
{\bf B}_2 {\bf B}_1 = \left[ \matrix{1 + (x_2 / x_1) \phi_2 & 
x_1 \phi_1 + x_2 \phi_2 \cr {1 / x_1} + {1 / x_2} & 
1 + (x_1 / x_2) \phi_1 } \right] \, . 
\ee
Thus, the product of two matrices contains only linear factors 
of $\phi_k$. As a result, we can define a new reference matrix 
${\bf \widetilde A}=\phi^{1/2}{\bf C}$ that
 accumulates factors of $\phi_k$ only half as
quickly as the original matrix $\bf A$ in the above argument, so that 
\be
{\bf \widetilde A}_2 {\bf \widetilde A}_1 = \phi_1^{1/2} \phi_2^{1/2} 
\left[ \matrix{1 + {x_2 / x_1} & x_1 + x_2 \cr 
{1 / x_1} + {1 / x_2} & 1 + {x_1 / x_2} } \right] \, = 
\phi_1^{1/2} \phi_2^{1/2} {\bf C}_2 {\bf C}_1 \, . 
\ee
The new reference matrix still grows more slowly than the matrix 
$\bf B$ of the full problem, but the product of $N$ such matrices 
accumulates only $N$ extra factors of $\phi_k^{1/2}$. Using this reference 
matrix in the above argument results in the tighter bound  
\be
0 < \varepsilon_{BC} < - {1 \over 2 \pi} 
\langle \log \fsubk \rangle \, . 
\label{eq:ebound} 
\ee 

In the limit where all of the $h_k \gg 1$, $\log \fsubk \approx
-1/h_k^2$, and the above bound approaches the approximate form
\be
0 < \varepsilon_{BC} < {1 \over 2 \pi} \langle h_k^{-2} \rangle \, . 
\ee 
This expression shows that the errors are well controlled. For large
but finite $h_k$, the departure of the growth rates from those
obtained in the highly unstable limit (Theorem 2) are 
${\cal O} (h_k^{-2})$. 
 $\thmbox$
Given the above considerations, we can write the growth rate 
$(\Delta \gamma)_B$ for the full problem in the form 
\be
(\Delta \gamma)_B = (\Delta \gamma)_C - 
{\coneps \over \pi} \langle h_k^{-2} \rangle \, , 
\ee
where $(\Delta \gamma)_C$ is the growth rate for the reduced problem
and where $\coneps$ is a constant of order unity. In the limit of
large $h_k$ (specifically, for $\log \fsubk \approx 1/h_k^2$), the
constant is bounded and lies in the range $0 < \coneps < 1/2$. Our
numerical exploration of parameter space suggest that $\coneps \approx
1/4$ provides a good estimate for the correction term. In any case, 
however, the correction term depends on $h_k$ through the quantity 
$\langle h_k^{-2} \rangle$ and decreases with the size of this 
expectation value. 

\subsection{Matrix Elements with Varying Signs} 

We now consider the case in which the signs of the variables $r_j$ can
be either positive or negative.  Suppose that the system has equal
probability of attaining positive and negative factors. In the limit
$N \to \infty$, one expects the sums $\sumtn, \sumbn \to 0$, which
would seem to imply no growth. However, two effects counteract this
tendency. First, the other factor that arises in the repeated matrix
multiplication diverges in the same limit, i.e.,
\be
\prod_k^N (2 h_k) \to \infty \qquad {\rm as} \qquad 
N \to \infty \, . 
\label{eq:diverge} 
\ee 
Second, the sums $\sumtn$ and $\sumbn$ can random walk away from zero
with increasing number $N$ of cycles, where the effective step length
is determined by the variance $\sigma_0$ defined previously. If the
random walk is fast enough, the system can be unstable even without
considering the diverging product of equation (\ref{eq:diverge}). In
order to determine the stability (or instability) of the Hill's
equation in this case, we must thus determine how the sums $\sumtn$
and $\sumbn$ behave with increasing $N$.

\bigskip
\noindent
{\bf Theorem 4:} {\it Consider the case of Hill's equation in the unstable
limit with both positive and negative signs for the matrix
elements. Let positive signs occur with probability $p$ and negative
signs occur with probability $1 - p$. Then the general form of the
growth rate is given by}  
\be 
\Delta \gamma = \lim_{N \to \infty} {1 \over \pi N} \left\{
\bigl[ p^2 + (1-p)^2 \bigr] \sum_{j=1}^N \log \bigl(1 + 
\big| {x_{j1} \over x_{j2}} \big| \bigr)  + 2 p (1-p) \sum_{k=1}^N 
\log \Big| 1 - \big| {x_{k1} \over x_{k2}} \bigr| \Big| \right\} 
- {\log 2 \over \pi} \, . 
\label{eq:pmgrowth} 
\ee

\noindent 
{\bf Proof:} The same arguments leading to equation (\ref{eq:almost})
in the proof of Theorem 2 can be used, where the signs of the ratios
$x_{j1}/x_{j2}$ must be taken into account. If $p$ is the probability
of the $x_j$ variables being positive, the probability of the ratio of
two variables being positive will be given by $p^2 + (1 - p)^2$, i.e.,
the probability of getting either two positive signs or two negative
signs.  The probability of the ratio being negative is then $2p(1-p)$.
With this consideration of signs, the intermediate form of equation 
(\ref{eq:almost}) is modified to take the form 
\be 
\Delta \gamma + {\log 2 \over \pi} \approx {1 \over N \pi} 
\sum_{j=1}^{N_P} \log \left[ 1 + {|x_{j1}/x_{j2}| \sumbn + 
|x_{j2}/x_{j1}| \sumtn \over \sumbn + \sumtn } \right] \, 
\ee
$$\, \qquad \qquad \qquad + {1 \over N \pi} 
\sum_{j=1}^{N_Q} \log \left[ 1 - {|x_{j1}/x_{j2}| \sumbn + 
|x_{j2}/x_{j1}| \sumtn \over \sumbn + \sumtn } \right] \, , $$
where $N_P$ is the number of terms where the ratios have positive
signs and $N_Q$ is the number of terms where the ratios have negative
signs. In the limit $N \to \infty$, we argue (as before) that the sums 
$\sumbn$ and $\sumtn$ approach the same value. Notice also that the 
two sums can be either positive or negative, but they will both have
the same sign (by construction). As a result, we can divide the sums
out of the expression as before. In the limit $N \to \infty$, the
fraction $N_P/N \to p^2 + (1-p)^2$ and the fraction $N_Q/N \to 2 p
(1-p)$. After some rearrangement, we obtain the form of equation
(\ref{eq:pmgrowth}). $\thmbox$ 

\begin{figure}
\figurenum{2}
{\centerline{\epsscale{0.90} \plotone{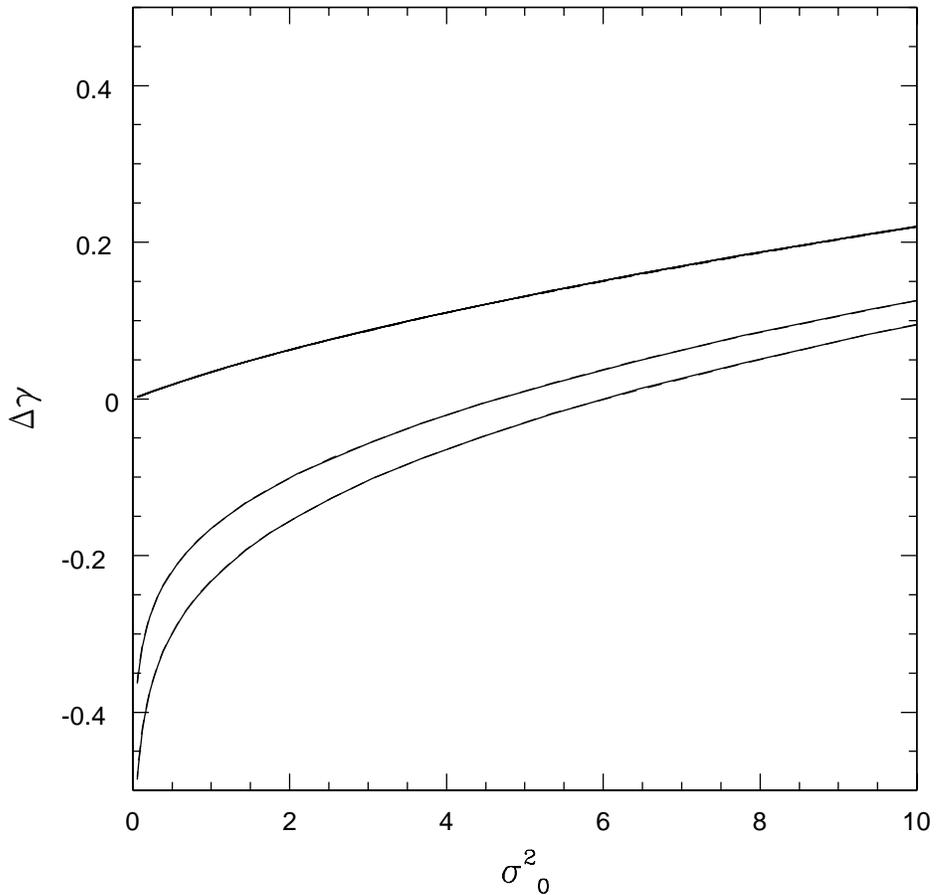} }} 
\figcaption{ Correction $\Delta \gamma$ to the growth rate for the
case in which the signs of the random variables $x_k$ are both
positive and negative. The three curves show the results for a 50/50
distribution (bottom), 75/25 (center), and the case of all positive
signs (top).  For all three cases, the solid curves show the results
of numerical matrix multiplication, where 1000 realizations of each
product are averaged together. The overlying dashed curves, which are
virtually indistinguishable, show the exact results from Theorem 4. }
\label{fig:mixed} 
\end{figure}

\begin{figure}
\figurenum{3}
{\centerline{\epsscale{0.90} \plotone{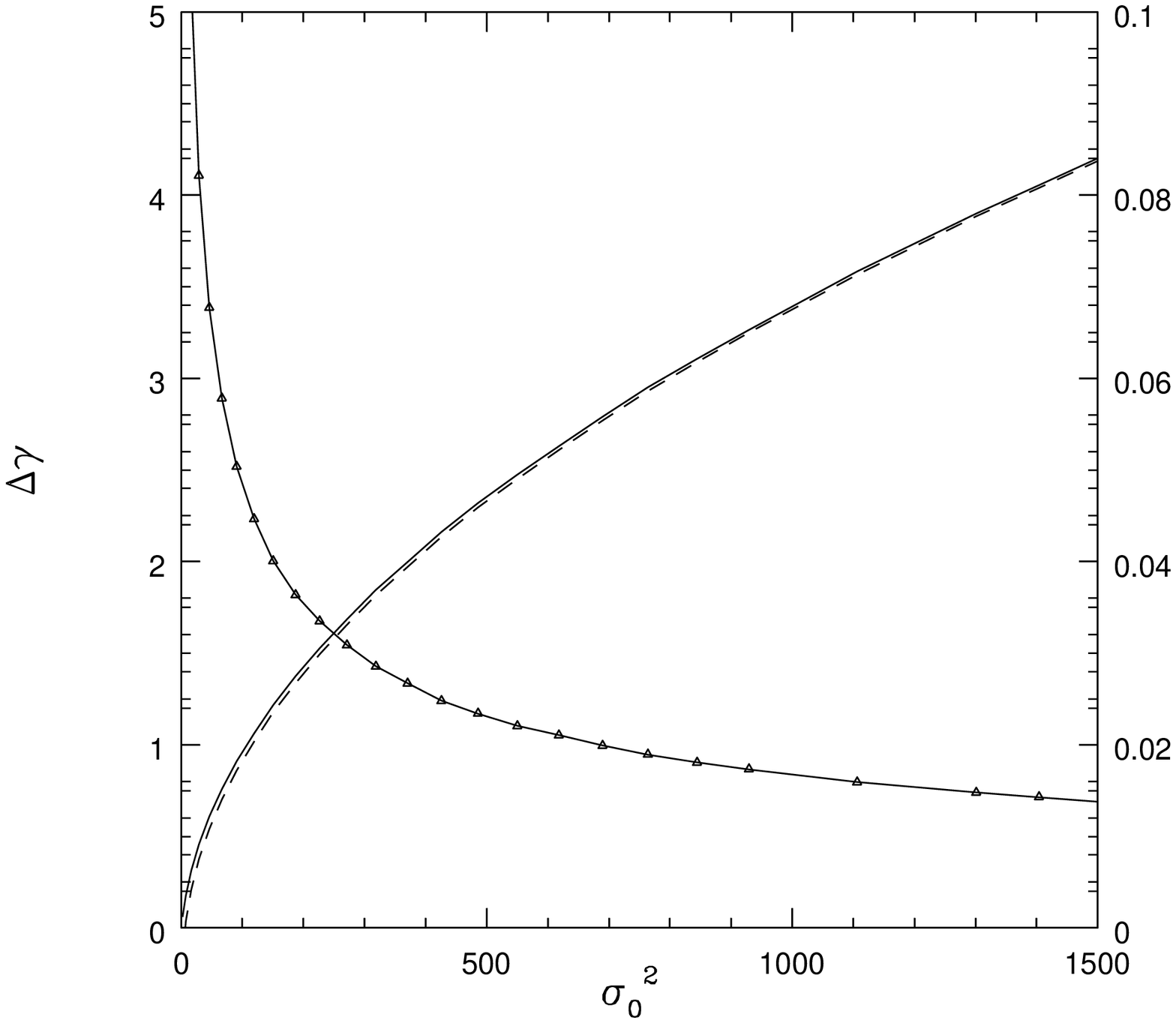} }} 
\figcaption{ Convergence of growth rates in the limit of large variance. 
The increasing solid curve shows the growth rate as a function of 
variance for the case of all positive signs.  The dashed curve shows the 
growth rate for the cased of mixed signs with a 50/50 sign distribution, 
i.e., $p$ = 1/2. The decreasing curve marked by triangles shows the 
difference between the two curves (where the axis on the right applies). } 
\label{fig:bigsigma} 
\end{figure}

\noindent 
{\bf Corollary 4.1:} {\it Let $P(\xi)$ denote the probability distribution
of the composite variable $\xi = x_k/x_j$, and assume that the
integral $\int d\xi (d P /d\xi) \log |\xi|$ exists.  Then for Hill's
equation in the unstable limit, and for the case of the variables $x_k$ 
having mixed signs, in the limit of small variance the correction to
the growth rate $\Delta \gamma$ approaches the following limiting form:} 
\be 
\lim_{\sigma_0 \to 0} \Delta \gamma = {2 p (1-p) \over \pi} 
\left[ \log \sigma_0 + \czero - \log 2 \right] \, , 
\ee 
{\it where $\czero$ is a constant that depends on the probability 
distribution of the variables $x_k$. } 

\noindent 
{\bf Proof:} In the limit of small $\sigma_0$, the variables $x_k$ can
be written in the form $x_k = 1 + \delta_k$ where $|\delta_k| \ll 1$.
To leading order, the expression of equation (\ref{eq:pmgrowth}) for 
the growth rate becomes 
\be 
\Delta \gamma + {\log 2 \over \pi} = \lim_{N \to \infty} 
{1 \over \pi N} \left\{ [p^2 + (p-1)^2] \sum_{j=1}^N 
\log ( 2 + \delta_{j1} - \delta_{j2} ) + 2 p (1 - p) 
\sum_{k=1}^N \log \bigl| \delta_{k1} - \delta_{k2} \bigr| 
\right\} \, . 
\ee 
In the limit of small variance $\sigma_0 \to 0$, the variables 
$\delta_k \to 0$, and the above expression reduces to the form 
\be 
\Delta \gamma = {2 p (1-p) \over \pi} \left[ \langle \log 
| \delta_{k1} - \delta_{k2} | \rangle - \log 2 \right] \, . 
\ee
We thus need to evaluate the expectation value given by 
\be
\langle \log |\delta_k - \delta_j| \rangle = 
\int d \xi \log |\xi| {d P \over d \xi} \, , 
\ee
where we have defined the composite variable $\xi = \delta_k -
\delta_j$. Notice that in the limit $|\delta| \ll 1$, the variance 
of $\xi$ is $\sigma_0^2$.  Next we define a dimensionless variable  
$z \equiv \xi/ \sigma_0$, so that the integral becomes 
\be
I = \int dz {d P \over dz} \log (\sigma_0 z) = \log \sigma_0 
\int dz {d P \over dz} + \int dz {d P \over dz} \log z = 
\log \sigma_0 + \int dz {d P \over dz} \log z \, . 
\ee 
As long as the differential probability distribution $dP/dz$ allows
the integral in the final expression to converge, then $I = \log
\sigma_0 + \czero$, where $\czero$ is some fixed number that depends
only on the shape of the probability distribution. This convergence
requirement is given by the statement of the corollary, so that
Corollary 4.1 holds. Notice also that in the limit of small
$\sigma_0$, the $\log \sigma_0$ term dominates for any fixed
$\czero$, so that $\Delta \gamma \sim 2p (1-p) (\log \sigma_0)/\pi$.
$\thmbox$

Figure \ref{fig:mixed} shows the growth rates as a function of the
variance $\sigma_0$ for the case of mixed signs. For the case of
positive signs only, $p$ = 1, the correction $\Delta \gamma$ to the
growth rate goes to zero as $\sigma_0 \to 0$. For the case of mixed
signs, the correction to the growth rate has the form $\Delta \gamma
\propto \log \sigma_0$ as implied by Corollary 4.1. 

Sometimes it is useful to explicitly denote when the growth rates
under consideration are the result of purely positive signs or mixed
signs for the variables $x_k$.  Here, we use the notation $\Delta
\gamma_p$ to specify the growth rate when all the signs are positive.
Similarly, $\Delta \gamma_q$ denotes growth rates for the case of
mixed signs.

\noindent
{\bf Corollary 4.2:} {\it In the limit of large variance, $\sigma_0 \to
\infty$, the growth rates for the case of positive signs only and for
the case of mixed signs converge, i.e., } 
\be 
\lim_{\sigma_0 \to \infty} \Delta \gamma_q = \Delta \gamma_p \, , 
\label{eq:cor42} 
\ee
{\it where $\Delta \gamma_p$ denotes the case of all positive signs
and $\Delta \gamma_q$ denotes the case of mixed signs. }

\noindent
{\bf Proof:} The difference in the growth rates for 
two cases is given by 
\be 
\Delta \gamma_p - \Delta \gamma_q = 
{2 p (1 - p) \over \pi} \lim_{N \to \infty} {1 \over N} 
\sum_{j=1}^N \left[ \log ( 1 + | x_{j1}/x_{j2} | ) - 
\log \Bigl| 1 - | x_{j1}/x_{j2} | \Bigr| \right] \, ,  
\label{eq:differ} 
\ee 
where $p$ is the probability for the sign of $x_k$ being positive.  
In the limit of large variance $\sigma_0^2 \to \infty$, the ratios
$|x_j/x_k|$ are almost always far from unity. Only the cases with
$|x_j/x_k| \gg 1$ have a significant contribution to the sums.  For
those cases, however, both of the logarithms in the sums reduce to the
same form, $\log |x_j/x_k|$, and hence equation (\ref{eq:differ}) 
becomes 
\be
\lim_{\sigma_0 \to \infty}
\Delta \gamma_p - \Delta \gamma_q = {2 p (1 - p) \over \pi} 
\left[ \langle \log |x_j/x_k| \rangle - \langle \log |x_j/x_k| 
\rangle \right] \to 0 \, . 
\ee 
As a result, equation (\ref{eq:cor42}) is valid. $\thmbox$ 

\noindent
{\bf Corollary 4.3:} {\it In the limit of large variance $\sigma_0 \to
\infty$, the growth rate $\Delta \gamma$ approaches the form given by}
\be
\lim_{\sigma_0 \to \infty} \Delta \gamma = {\sigma_0 \over \pi} \cbig \, , 
\ee 
{\it where $\cbig$ is a constant that depends on the form of the
probability distribution for the variables $x_k$. In general, 
$\cbig \le 1/2$. } 

\noindent 
{\bf Proof:} Let the composite variable $\xi = \log (x_k / x_j)$ have a
probability distribution $dP/d\xi$. Since the growth rate for the case
of mixed signs converges to that for all positive signs in the limit
of interest (from Corollary 4.2), we only need to consider the latter
case (from Theorem 2). The growth rate is then given by the
expectation value
\be 
\Delta \gamma = {1 \over \pi} \int_{-\infty}^\infty d\xi 
{dP \over d\xi} \log (1 + {\rm e}^\xi ) \, . 
\ee 
The integral can be separated into the domains $\xi < 0$ and 
$\xi > 0$.  For the positive integral, we expand the integrand into
two terms; for the negative domain, we change variables of integration
so that $\xi \to - \xi$.  We thus obtain the three terms
\be 
\Delta \gamma = {1 \over \pi} \int_0^\infty d\xi {dP \over d\xi} 
\xi \, + \, {1 \over \pi} \int_0^\infty d\xi {dP \over d\xi} 
\log (1 + {\rm e}^{-\xi} ) \, + \, {1 \over \pi} \int_0^\infty 
d\xi {d {\widetilde P} \over d\xi} \log (1 + {\rm e}^{-\xi} ) \, . 
\ee 
In the third integral, the probability distribution 
$(d{\widetilde P}/d\xi)(\xi)$ = $(dP/d\xi) (-\xi)$; the second and
third terms will thus be the same since the distribution is symmetric
(by construction, the composite variable $\xi$ is the difference 
between two variables $\log x_k$ drawn from the same distribution). 
The sum of the second two integrals is bounded from above by $\log 2$
and can be neglected in the limit of interest. In the first integral,
we change variables according to $z = \xi/\sigma$, so that
\be 
\Delta \gamma \to {\sigma_0 \over \pi} \langle z \rangle_{(\xi \ge 0)}  
\qquad {\rm where} \qquad \langle z \rangle_{(\xi \ge 0)}   
\equiv \int_0^\infty dz {dP \over dz} z \, . 
\ee  
Since $\langle 1 \rangle$ = 1 and $\langle z^2 \rangle$ = 1, by
definition, we expect the quantity $\langle z \rangle_{(\xi \ge 0)}$ 
= $\cbig$ to be of order unity. Further, one can show that $\cbig$ 
as defined here is bounded from above by 1/2. As a result, in this 
limit, we obtain a bound of the form $\pi (\Delta \gamma) \le$
$\sigma_0 / 2 + \log 2$. We note that the constant $\cbig$ cannot 
be bounded from below (in the absence of further constraints 
placed on the probability distribution $dP/d\xi$).  $\thmbox$ 

\noindent
{\bf Corollary 4.4:} {\it In the limit of large variance $\sigma_0^2 \gg 1$,
the difference $\Delta (\Delta \gamma)$ between the growth rate for
strictly positive signs and that for mixed signs takes the form} 
\be
\lim_{\sigma_0 \to \infty} \Delta (\Delta \gamma) = 
{8 p (1 - p) \over \pi \sigma_0} \cdelt \, , 
\ee
{\it where $\cdelt$ is a constant that depends on the form of 
probability distribution, and where $p$ is the probability 
of positive matrix elements for the case of mixed signs. }

\noindent
{\bf Proof:} Using the results from Theorem 2 and Theorem 4 to specify
the growth rates for the cases of positive signs and mixed signs,
respectively, the difference can be written in the form
\be 
\Delta (\Delta \gamma) = {2 p (1-p) \over \pi} 
\int_{-\infty}^\infty d\xi \, {d P \over d \xi} 
\left[ \log (1 + {\rm e}^\xi ) - 
\log \big| 1 - {\rm e}^\xi \big| \right] \, . 
\ee
Next we separate the integrals into positive and negative 
domains and change the integration variable for the negative 
domain ($\xi \to - \xi$). The integral ($I$) then becomes 
\be
I = \int_0^\infty d\xi {d P \over d\xi} \log \left( 
{1 + {\rm e}^{-\xi} \over 1 - {\rm e}^{-\xi} } \right) 
+ \int_0^\infty d\xi {d {\widetilde P} \over d\xi} 
\left( {1 + {\rm e}^{-\xi} \over 1 - {\rm e}^{-\xi} } \right) \, , 
\ee
where ${\widetilde P} (\xi) = P(-\xi)$. 
Since we are working in the large $\sigma_0$ limit, 
the variable $\xi$ will be large over most of the domain 
where the integrals have support, so we can expand using 
${\rm e}^{-\xi}$ as a small parameter. In this case, the 
integral $I$ becomes 
\be 
I = 2 \int_{-\infty}^\infty d \xi \, {d P \over d \xi} 
{\rm e}^{-| \xi | } \, = \, 2 \int_{-\infty}^\infty 
dz {d P \over dz}  {\rm e}^{- \sigma_0 |z| } \, , 
\ee
where we have made the substitution $z = \xi/\sigma$.  For large
$\sigma_0$, the decaying exponential dominates the behavior of the
integrand. In the limit $\sigma \to \infty$, the exponential term
decays to zero before the probability $dP/dz$ changes so that 
$dP/dz \to \cdelt$, where $\cdelt$ is a constant. The integral thus
becomes $I = 4 \cdelt / \sigma_0$, and the difference between the
growth rates becomes
\be 
\Delta (\Delta \gamma) = {8 p (1 - p) \over \pi \sigma_0} 
\cdelt \, , 
\ee
as claimed by Corollary 4.4. $\thmbox$ 

Figure \ref{fig:bigsigma} illustrates the behavior implied by the last
three Corollaries.  In the limit of large variance, the growth rates
for mixed signs and positive signs only converge (Corollary
4.2). Further, growth rates for both cases approach the form $\Delta
\gamma \propto \sigma_0$ (as in Corollary 4.3). Finally, the
difference between the growth rates for the two cases has the
characteristic form $\Delta (\Delta \gamma) \propto 1/\sigma_0$ (from
Corollary 4.4). 

\noindent
{\bf Corollary 4.5:} {\it For the case of mixed signs, the crossover 
point between growing solutions and decaying solutions is given by 
the condition }
\be 
[p^2 + (1-p)^2 ] \langle \log \big| 1 + |x_j / x_k| \big| \rangle + 
2 p (1-p) \langle \log \big| 1 - |x_j / x_k | \big| \rangle \, = \log 2 . 
\label{eq:crossover}
\ee

\noindent
{\bf Proof:} This result follows form Theorem 4 by inspection. 
$\thmbox$ 

\noindent 
{\sl Estimate for the Crossover Condition:} Equation
(\ref{eq:crossover}) is difficult to evaluate in practice. In order to
obtain a rough estimate of the threshold for instability, we can
consider the the $r_j$ to be independent variables and use elementary
methods to estimate the conditions necessary for systems with mixed
signs to be unstable.  We first note that the sums $\sumtn$ and
$\sumbn$ add up the composite variables $r_j$, which are made up of
the variables $x_j$ (which in turn are set by the form of the original
differential equation). If the signs of the variables $x_j$ are
symmetrically distributed, then the signs of the composite variables
$r_j$ are also symmetrically distributed. We can thus focus on the
variables $r_j$.

Since the signs can either be positive or negative, the probability of
a net excess of positive (or negative) terms is governed by the
binomial distribution (which has a gaussian form in the limit of large
$N$).  The probability $P$ of having a net excess of $m$ signs is
given by the distribution
\be
P(m) =  (\pi N_S / 2)^{-1/2} 
\exp \bigl[ - m^2 / 2 N_S \bigr] \, , 
\ee
where $N_S$ is the number of steps in the random walk. 
The sums $\sumtn$ and $\sumbn$ have $N_S = 2^N$ steps, 
where $N$ is the number of cycles of the Hill's equation. 

If the net excess of signs of one type is $m$, the sums are reduced
(from those obtained with purely positive variables) so that
\be
\swig = \swig_0 {m \over N_S} \, , 
\ee
where $\swig_0$ is the value of the composite sum obtained 
when the variables $x_j$ have only one sign. 

The probability of a growing solution is given by 
\be
P_G = \int_{m_\ast}^\infty P(m) dm \, , 
\ee
where $m_\ast$ is the minimum number of steps needed 
for instability. We can write $m_\ast$ in the form 
\be
m_\ast = N_S {\rm e}^{-N \pi \dgamz } = \exp  
\bigl[ N (\log(2) - \pi \dgamz ) \bigr] \, ,  
\ee
where $\dgamz$ is the correction to the growth rate 
for the case of positive signs only.

The integral can be written in terms of the variable 
$\xi = m / (2N_S)^{1/2}$ so that 
\be 
P_G = {2 \over \sqrt{\pi} } 
\int_{z_\ast}^\infty {\rm e}^{-z^2} dz \, , 
\ee 
where 
\be
z_\ast = \exp \bigl[ N \bigl( 
{1 \over 2} \log 2 - \pi \dgamz \bigr] \, . 
\ee
Thus, the crossover for growth occurs under the condition 
\be 
\dgamz  \approx \log2/(2 \pi) \  . 
\ee
Keep in mind that this result was derived under the assumption that
the variables in the random walk are completely independent.  We can
derive the above approximate result from a simpler argument: The sums
$\sumtn$ and $\sumbn$ random walk away from zero according to $\ell
\sqrt{N_S} = \langle r_j^2 \rangle^{1/2} 2^{N/2} = \exp [ n \sigma_0^2
+ (N/2) \log 2]$.  As a result, $\swig \approx \exp [n \sigma_0^2 -
(N/2) \log2]$ and hence $\Delta \gamma \approx (n/N) (\sigma_0^2/\pi)
- (\log 2)/ 2 \pi$.

\subsection{Specific Results for a Normal Distribution} 

In this section we consider the particular case where the composite
variable $\xi = \log (x_k/x_j)$ has a normal distribution. Specifically, 
we let the differential probability distribution take the form 
\be 
{d P \over d \xi} = {1 \over \sqrt{2 \pi} \sigma_0} 
{\rm e}^{-\xi^2/2 \sigma_0^2} \, , 
\ee 
so that $\sigma_0^2$ is the variance of the distribution. In order to 
determine the growth rates, we must evaluate the integrals
\be
J_\pm = {1 \over \sqrt{2 \pi} \sigma_0} \int_{-\infty}^\infty d\xi \, 
{\rm e}^{-\xi^2/2 \sigma_0^2} \, \log \left| 1 \pm {\rm e}^\xi \right| \, . 
\ee 

In the limit $\sigma_0 \to 0$, the correction part of the 
growth rate ($\Delta \gamma$) can be evaluated and has the form 
\be 
\lim_{\sigma_0 \to 0} \Delta \gamma = {1 \over \pi} 
\left\{ \bigl[ p^2 + (1-p)^2 \bigr] {\sigma_0^2 \over 8} + 
2 p (1-p) \bigl[ \log \sigma_0 - {\gem \over 2} \bigr] - 
3 p (1-p) \log 2 \right\} \, , 
\ee 
where $\gem = 0.577215665\dots$ is the Euler-Mascheroni constant. Note
that for the case of positive signs only ($p=1$), this expression
reduces to the form $\Delta \gamma = \sigma_0^2/ (8 \pi)$ as in
Corollary 2.1.  For the case of mixed signs, this expression reduces
to the form $\Delta \gamma \propto \log \sigma_0$ from Corollary 4.1.

We can also evaluate the growth rate in the limit of large $\sigma_0$, 
and find the asymptotic form 
\be 
\lim_{\sigma_0 \to \infty} \Delta \gamma = 
{\sigma_0 \over \sqrt{2} \pi^{3/2}} \, . 
\ee 
As a result, the constant $\cbig$ from Corollary 4.3 is given by
$\cbig = 1/\sqrt{2\pi}$. Note that in this limit, the growth rate is
independent of the probabilities $p$ and $(1-p)$ for the variables
$x_k$ to have positive and negative signs, consistent with Corollary
4.2.  In this limit, we can also evaluate the difference between the
case of positive signs and mixed signs, i.e.,
\be 
\Delta \gamma_p - \Delta \gamma_q = {8 p (1-p) \over 
\sqrt{2} \pi^{3/2} \sigma_0 } \, . 
\ee 
Thus, the constant $\cdelt$ from Corollary 4.4 is given by 
$\cdelt = 1 / \sqrt{2 \pi}$ for the case of a normal distribution. 
Note that although $\cdelt = \cbig$ for this particular example, 
these constants will not be the same in general. 

Finally, for the case of purely positive signs, we can connect the
limiting forms for small variance and large variance to construct a
rough approximation for the whole range of $\sigma_0$, i.e., 
\be
\Delta \gamma \approx 
{\sigma_0^2 / \pi \over 8 + \sqrt{2 \pi} \sigma_0} \, .
\ee 
This simple expression, which is exact in the limits $\sigma_0 \to 0$
and $\sigma_0 \to \infty$, has a maximum error of about 18\% over
the entire range of $\sigma_0$.

\subsection{Matrix Decomposition for Small Variance} 

For completeness, and as a consistency check, we can study the growth
rates by breaking the transformation matrix into separate parts. In
this section we consider the case of small variance (see Appendix B 
for an alternate, more general, separation).  In the limit of small
variance, $\sigma_0^2 \ll 1$, the variables $x_k$ only have small
departures from unity and can be written in the form
\be 
x_k = 1 + \delta_k \, , 
\ee 
where $|\delta_k| \ll 1$.  The matrices of the discrete map can then
be decomposed into two parts so that
\be 
{\bf C}_k = {\bf A}_k + s_k \delta_k {\bf B}_k \, , 
\label{eq:cksum} 
\ee
where $s_k = \pm 1$ is the sign of the $k$th term, and where  
\be 
{\bf A}_k = \left[ \matrix{1 & s_k \cr s_k & 1 } \right] 
\qquad {\rm and} \qquad 
{\bf B}_k = \left[ \matrix{0 & 1 \cr -1 & 0 } \right] \, . 
\ee 
The matrices ${\bf A}_k$ and ${\bf B}_k$ have simple multiplicative
properties. In particular, 
\be 
{\bf A}_j {\bf A}_k = 2 {\bf A}_j \quad {\rm if} \, \, s_j = s_k \, , 
\qquad {\rm but} \qquad {\bf A}_j {\bf A}_k = 0 
\quad {\rm if} \, \, s_j \ne s_k \, , 
\ee 
and 
\be 
{\bf B}_k^2 = - {\bf I} \, , \qquad  
{\bf B}_k^3 = - {\bf B}_k \, , \qquad  {\rm and} \qquad 
{\bf B}_k^4 = {\bf I} \, .
\ee 
The product matrix $\prod {\bf C}_k$ will contain long strings of
matrices ${\bf A}_k$ and ${\bf B}_k$ multiplied by each other. If any
two matrices ${\bf A}_k$ have opposite signs in such a multiplication
string, then the product of the two matrices will be zero and the
entire string will vanish. As a result, after a large number $N$ of
cycles, the only matrices that are guaranteed to survive in the
product are those with only ${\bf B}_k$ factors and those with only
one ${\bf A}_k$ factor. Although it is possible for strings with
larger numbers of ${\bf A}_k$ to survive, it becomes increasingly
unlikely (exponentially) as the number of factors increases. To a good
approximation, the eigenvalue of the resulting product matrix will be
given by the product
\be
\Lambda^{(N)} \approx \prod_{k=1}^N \delta_k \, . 
\ee 
We could correct for the possibility of longer surviving strings of 
${\bf A}_k$ by multiplying by a factor of order unity; however, such 
a factor would have a vanishing contribution to the growth rate. 
The corresponding growth rate thus takes the form 
\be
\Delta \gamma = \lim_{N \to \infty} {2 p(1-p) \over N \pi} 
\sum_{k=1}^N \log |\delta_k| \, , 
\ee 
where the factor $2p(1-p)$ arises because the matrices with 
all positive signs lead to a zero growth rate in the limit 
$\sigma_0 \to 0$, so only the fraction of the cases with mixed 
signs contribute. Next we note that the sum converges to an  
expectation value 
\be
\langle |\delta_k| \rangle = \int d \delta {dP \over d \delta} 
\log|\delta| \, . 
\ee 
Next we make the substitution $z = \delta/\sigma_0$ and 
rewrite the integral in the form 
\be
\langle |\delta_k| \rangle = \sigma_0 \int dz {dP \over dz} + 
\int dz {dP \over dz} \log z \, . 
\ee 
In the limit of interest, $\sigma_0 \to 0$, the first term 
dominates and the growth rate (to leading order) approaches 
the form 
\be 
\Delta \gamma = {2 p(1-p) \over \pi} \log \sigma_0 \, . 
\ee 
This form agrees with the leading order expression found earlier in
Corollary 4.1 (see also Figure \ref{fig:mixed}, which shows the growth
rate as a function of the variance).

\section{HILL'S EQUATION IN THE DELTA FUNCTION LIMIT} 

In many physical applications, including the astrophysical orbit
problem that motivated this analysis, we can consider the forcing
potential to be sufficiently sharp so that $\qhat (t)$ can be
considered as a Dirac delta function. For this limit, we specify the
main equation considered in this section:

\noindent 
{\bf Definition:} {\it Hill's equation in the delta function limit} 
is defined to have the form 
\be 
{d^2 y \over dt^2} + [ \af + q \delta (\tmod - \pi/2) ] y = 0 \, ,
\label{eq:eqdelta} 
\ee 
where $q$ measures the strength of the forcing potential and where
$\delta (t)$ is the Dirac delta function. In this form, the time
variable is scaled so that the period of one cycle is $\pi$.  The
argument of the delta function is written in terms of $\tmod$, which
corresponds to the time variable mod-$\pi$, so that the forcing
potential is $\pi$-periodic.

This form of Hill's equation allows for analytic solutions, as
outlined below, which can be used to further elucidate the instability
for random Hill's equations.  In particular, in this case, we can
solve for the transformation between the variables $(\af_k, q_k)$ that
appear in Hill's equation and the derived composite variables $x_k$
that determine the growth rates.

\subsection{Principal Solutions} 

To start the analysis, we first construct the principal solutions to
equation (\ref{eq:eqdelta}) for a particular cycle with given values
of forcing strength $q$ and oscillation parameter $\af$. The equation
has two linearly independent solutions $y_1(t)$ and $y_2(t)$, which
are defined through their initial conditions
\be 
y_1 (0) = 1, \quad {dy_1 \over dt} (0)   = 0 , \qquad {\rm and} \qquad 
y_2 (0) = 0, \quad {dy_2 \over dt} (0) = 1 \, . 
\label{eq:bc} 
\ee 
The first solution $y_1$ has the generic form 
\be 
y_1 (t) = \cos \sqrt{\af} t 
\qquad {\rm for} \quad 0 \le t < \pi/ 2 \, , 
\ee
and 
\be
y_1 (t) = A \cos \sqrt{\af} t + B \sin \sqrt{\af} t 
\qquad {\rm for} \quad \pi/ 2 < t \le \pi \, , 
\ee
where $A$ and $B$ are constants that are determined by  
matching the solutions across the delta function at $t = \pi/2$. 
We define $\theta \equiv \sqrt{\af} \pi/2$ and find 
\be 
A = 1 + (q/\sqrt{\af}) \sin\theta \cos\theta 
\qquad {\rm and} \qquad 
B = - (q/\sqrt{\af}) \cos^2 \theta \, . 
\ee
Similarly, the second solution $y_2$ has the form 
\be
y_2 (t) = \sin \sqrt{\af} t \qquad 
{\rm for} \quad 0 < t < \pi/2 \, , 
\ee
and 
\be 
y_2 (t) = C \cos \sqrt{\af} t + D \sin \sqrt{\af} t 
\qquad {\rm for} \quad \pi/ 2 < t \le \pi \, , 
\ee
where 
\be 
C = (q/\af) \sin^2 \theta \qquad {\rm and} \qquad 
D = {1 \over \sqrt{\af} } - (q/\af) \sin\theta \cos\theta 
\, . 
\ee 

\begin{figure}
\figurenum{4}
{\centerline{\epsscale{0.90} \plotone{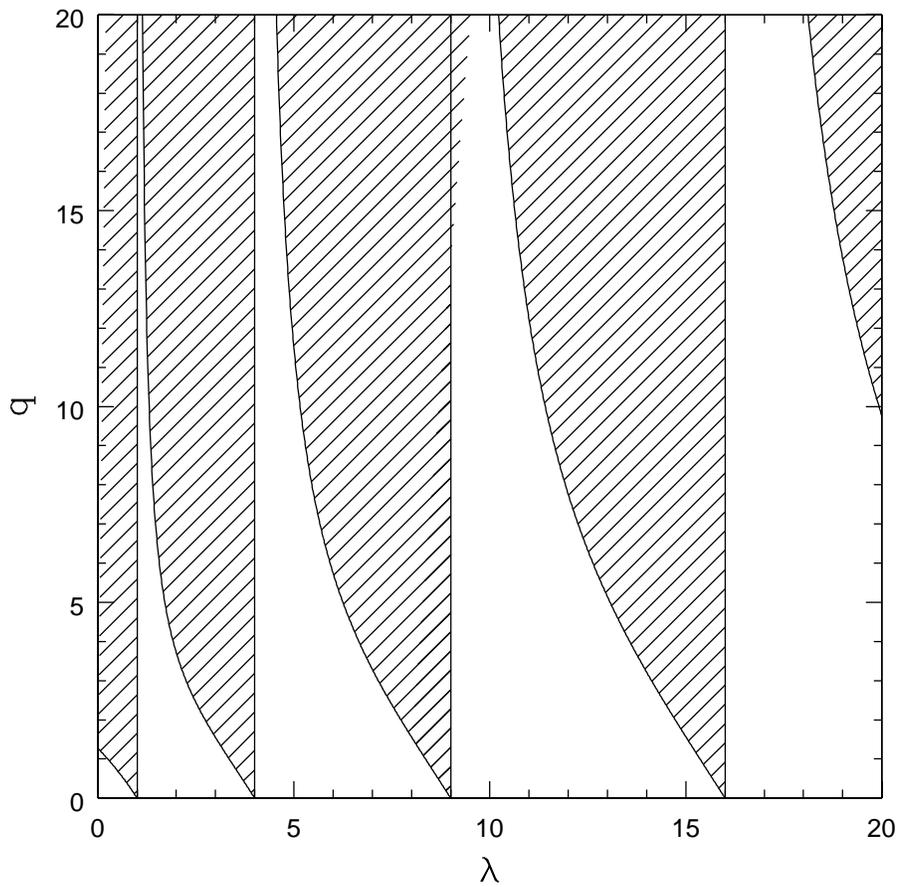} }} 
\figcaption{ Regions of instability for Hill's equation in the delta
function limit.  The shaded regions show the values of $(\af,q)$
that correspond to exponentially growing (unstable) solutions, which 
represent unstable growth of the perpendicular coordinate for orbits 
in our triaxial potential that are initial confined to one of the 
principal planes. } 
\label{fig:qlplane} 
\end{figure}

For the case of constant parameters $(q,\af)$, we can find the
criterion for instability and the growth rate for unstable solutions.
Since the forcing potential is symmetric, $y_1 (\pi) = dy_2/dt (\pi)$,
from Theorem 1.1 of [MW]. The resulting criterion for instability
reduces to the form
\be 
H \equiv \Bigg| {q \over 2 \sqrt{\af} } \sin (\sqrt{\af} \pi) - 
\cos (\sqrt{\af} \pi) \Bigg| > 1 \, , 
\label{eq:hdef} 
\ee 
and the growth rate $\gamma$ is given by 
\be 
\gamma = {1 \over \pi} \log [ H + \sqrt{H^2 - 1} ] \, .
\ee
In the delta function limit, the solution to Hill's equation is thus 
specified by two parameters: the frequency parameter $\af$ and the
forcing strength $q$.  Figure \ref{fig:qlplane} shows the plane of
possible parameter space for Hill's equation in this limit, with the 
unstable regions shaded. Note that a large fraction of the plane is
unstable.

\subsection{Random Variations in the Forcing Strength} 

We now generalize to the case where the forcing strength $q$ varies
from cycle to cycle, but the oscillation parameter $\af$ is fixed.
This version of the problem describes orbits in triaxial, extended
mass distributions [AB] and is thus of interest in astrophysics.  As
outlined in \S 2.2, the solutions from cycle to cycle are connected by
the transformation matrix given by equation (\ref{eq:matrixdef}). Here, 
the matrix elements are given by 
\be
h = \cos (\sqrt{\af}\pi) - {q \over 2 \sqrt{\af} } 
\sin (\sqrt{\af}\pi) 
\qquad {\rm and} \qquad 
g = - \sqrt{\af} \sin (\sqrt{\af}\pi) - 
q \cos^2 (\sqrt{\af}\pi/2) \, . 
\label{eq:elements} 
\ee

\noindent 
{\bf Theorem 5:} {\it Consider a random Hill's equation in the delta
function limit. For the case of fixed $\lambda$, the growth rate of
instability approaches the asymptotic growth rate $\gamma_\infty$ in
the highly unstable limit $q/\sqrt{\lambda} \gg 1$, where the 
correction term has the following order:} 
\be 
\gamma \to \gamma_\infty \Biggl\{ 1 + 
{\cal O} \Bigl( {\lambda / q^2 } \Bigr) \Biggr\}  \, . 
\ee 

\noindent 
{\bf Corollary 5.1:} {\it In the delta function limit, the random Hill's
equation with fixed $\lambda$ is unstable when the asymptotic growth 
rate $\gamma_\infty > 0$.} 

\noindent
{\bf Remark 5.2:} {\it Note that $\gamma_\infty > 0$ requires only that a
non-vanishing fraction of the cycles are unstable.} 

\noindent
{\bf Proof:} For this version of the problem, the matrix $\bf M$
represents the transition from one cycle to the next, where the
solutions are written as linear combinations of $y_1$ and $y_2$ for
the given cycle.  In other words, this transformation operates in the
$(y_1, y_2)$ basis of solutions. However, one can also consider the
purely growing and decaying solutions, which we denote here as $\grow$
and $\decay$.

For a given cycle, the eigenvectors $V_{\pm}$ of the matrix $\bf M$
take the form 
\be
V_{\pm} = \left[ \matrix{ 1 \cr \pm g/k \cr } \right] \, , 
\ee
where the $+$($-$) sign refers to the growing (decaying) solution.
The eigenvalues have the form $\Lambda_{\pm}$ = $h \pm k$, where $k
\equiv (h^2 - 1)^{1/2}$. Keep in mind that $h = y_1 (\pi)$ and $g =
{\dot y}_2 (\pi)$, and that $\Lambda_{-} = 1/\Lambda_{+}$.  We can
write any general solution in the form
\be
f = A V_{+} + B V_{-} \, , 
\ee 
where the coefficients $(A,B)$ are related to the coefficients 
$(\alpha, \beta)$ in the first basis through the transformation 
\be 
\left[ \matrix{ A \cr B \cr } \right] \, =  {1 \over 2} 
\left[ \matrix{ 1 & k/g \cr 1 & -k/g \cr } \right] \, 
\left[ \matrix{ \alpha \cr \beta \cr } \right] \, . 
\label{eq:trans} 
\ee 
In the basis of eigenvectors, the action of the differential equation
over any cycle is to amplify growing solution (eigenvector) and
attenuate the decaying solution, and this action can be written as 
the matrix transformation
\be 
\left[ \matrix{ A' \cr B' \cr } \right] \, = 
\left[ \matrix{ \Lambda_{+} & 0 \cr 0 & \Lambda_{-} \cr } \right] \, 
\left[ \matrix{ A \cr B \cr } \right] \, . 
\ee 
At the end of the cycle, we can transform back to the original basis
through the inverse of the transformation (\ref{eq:trans}). As a
result, the original matrix $\bf M$ can be decomposed into three
components so that
\be 
{\bf M} (q, \af) = {1 \over 2} 
\left[ \matrix{ 1 & 1 \cr g/k & -g/k \cr } \right] \, 
\left[ \matrix{ \Lambda_{+} & 0 \cr 0 & \Lambda_{-} \cr } \right] \, 
\left[ \matrix{ 1 & k/g \cr 1 & -k/g \cr } \right] \, .
\ee
For each cycle, the values of $(q,\af)$ can vary. The next cycle
will have a new matrix of the same general form, with the matrix
elements specified by $(q',\af')$. 

We now shift our view to the basis of eigenvectors, so that each cycle
amplifies the growing solution.  Between the applications of the
amplification factors, the action of successive cycles ``rotates'' the
solution according to a transition matrix of the form
\be 
{\bf T} (q, \af; q', \af') = {1 \over 2} 
\left[ \matrix{ 1 & k'/g' \cr 1 & -k'/g' \cr } \right] \, .
\left[ \matrix{ 1 & 1 \cr g/k & -g/k \cr } \right] \, = {1 \over 2} 
\left[ \matrix{ 1 + \rat & 1 - \rat \cr  1 - \rat & 1 + \rat \cr } \right] \, , 
\ee
where the primes denote the second cycle and where we have defined
$\rat \equiv k'g/(k g')$. For the case in which successive cycles have
the same values of the original parameters $(q,\af)$, the
transition matrix $\bf T$ becomes the identity matrix (as expected).

For simplicity, we now specialize to the case where $\af$ is held
constant from cycle to cycle, but the forcing strength $q$ varies. We
can evaluate the transition matrix for the case in which Hill's
equation lies in the delta function limit and where we also take the
limit $q / \sqrt{\af} \gg 1$. In this regime,
\be 
\rat = 1 + {q - q' \over q'} \, {2 \sqrt{\af} \over q} \, \,  
{1 - 2 \cos (\sqrt{\af} \pi) \over \sin (\sqrt{\af} \pi) } 
+ {\cal O} \Bigl( {\lambda \over q^2 } \Bigr) \, \equiv 1 + 2 \delta . 
\ee   
Note that $\rat = 1 + 2 \delta$ to leading order, where $\delta$
(defined through the above relation) is small compared to unity and
the sign of $\delta$ can be both positive and negative. Thus, not only
is the parameter $\delta$ small, but it can average to zero. Repeated 
iterations of the mapping lead to the (1,1) matrix element growing 
according to the product 
\be 
M_{(1,1)} = \prod_{k=1}^N \left[ \Lambda_k (1 + \delta_k) \right] \, \approx 
\left[ \prod_{k=1}^N \Lambda_k \right] \, \left[ 1 + \sum_{k=1}^N \delta_k 
+ \sum_{k=1}^N {\cal O}(\delta_k^2) \right] \, . 
\ee 
The other matrix elements are of lower order (in powers of $1/q$) so
that to leading order the growing eigenvalue of the product matrix is
equal to the (1,1) matrix element. Further, for sufficiently
well-behaved distributions of the parameter $q$, the sum of $\delta_k$
averages to zero as $N \to \infty$. The growth rate is thus given by
\be 
\gamma = {1 \over \pi N} \sum_{k=1}^N \log(\Lambda_k) + 
{1 \over \pi N} \sum_{k=1}^N \log ( 1 + \delta_k) = 
\gamma_\infty + {\cal O} \Bigl( {\lambda \over q^2 } \Bigr) \, . 
\ee 
The condition required for the $\delta_k$ to average to zero 
can be expressed in the form 
\be 
\lim_{N \to \infty} {1\over N} \sum_{k=1}^N {q' - q \over q q'} 
= \langle {1 \over q} \rangle - \langle {1 \over q'} \rangle 
= 0 \, , 
\ee
which will hold provided that the expectation value 
$\langle 1/q \rangle$ exists. This constraint is nontrivial, 
in that a uniform probability distribution $P(q)$ = {\it constant} 
that extends to $q=0$ will produce a divergent expectation value 
for $\langle 1/q \rangle$. Fortunately, in the physical application 
that motivated this analysis, the value of $q$ is determined by 
the distance to the center of an orbit (appropriately weighted) 
so that the minimum value of $q$ corresponds to the maximum value 
of the distance. Since physical orbits have a maximum outer 
turning point (due to conservation of energy), physical orbit 
problems will satisfy the required constraint on the probability 
distribution. $\thmbox$ 

\subsection{Second Matrix Decomposition} 

Another way to decompose the transformation matrix is to separate it
into two separate rotations, one part that is independent of the
forcing strength $q$, and another that is proportional to $q$. We can
thus write the matrix in the form
\be 
{\bf M} (q, \af) = {\bf A} - {q \over 2 \sqrt{\af} } {\bf B} \equiv  
\left[ \matrix{ \cos 2\theta & (\sin 2 \theta)/\sqrt{\af} \cr 
- \sqrt{\af} \sin 2 \theta & \cos 2 \theta \cr } \right] \, - 
{q \over 2 \sqrt{\af} } 
\left[ \matrix{ \sin 2\theta & (2 \sin^2 \theta)/\sqrt{\af} \cr 
2 \sqrt{\af} \cos^2 \theta & \sin 2 \theta \cr } \right] \, , 
\label{eq:abdecomp} 
\ee
where the second equality defines the matrices $\bf A$ and $\bf B$.  
With these definitions, one finds that 
\be 
{\bf A}^N (\theta) = {\bf A} (N \theta) \qquad {\rm and} \qquad 
{\bf B}^N (\theta) =  ( 2 \sin 2 \theta)^{N-1} {\bf B} (\theta) \, , 
\ee 
where we again take $\lambda$ to be constant from cycle to cycle. 
As a result, after $N$ cycles, the effective transformation matrix 
can be written in the form 
\be 
{\bf M}^{(N)} = \prod_{k=1}^N \bigl({\bf A} - 
{q_k \over 2 \sqrt{\af}} {\bf B} \bigr) \, \, . 
\ee
In the asymptotic limit $q/\sqrt{\af} \to \infty$, 
the matrix approaches the form 
\be 
{\bf M}^{(N)} = \,  (-1)^N \, 
\left[ \prod_{k=1}^N {q_k \over 2 \sqrt{\af}} \right] 
\, \, (2 \sin 2 \theta)^{N-1} {\bf B} (\theta) \, . 
\ee 
The condition for stability takes the form 
$| {\rm Tr} {\bf M}^{(N)}| \ge 2$, i.e., 
\be 
\left[ \prod_{k=1}^N q_k \right] 
\, \, \left[ {\sin 2 \theta \over \sqrt{\af}} \right]^N  
\ge 1 \, . 
\ee 
When the system is unstable, the factor on the left hand side of
this equation represents the growth factor over the entire set of 
$N$ cycles. The growth rate $\gamma$ is thus given by 
\be 
\gamma = \lim_{N \to \infty} {1 \over \pi N} 
\log \left[ \prod_{k=1}^N \Bigl( q_k 
{\sin 2 \theta \over \sqrt{\af}} \Bigr) \right] \, = 
\lim_{N \to \infty} {1 \over \pi N} \sum_{k=1}^N \log 
\Bigl( q_k {\sin 2 \theta \over \sqrt{\af}} \Bigr) \, . 
\ee
Since $H_k = q_k (\sin 2 \theta) / \sqrt{\af}$ in this 
asymptotic limit, the above expression for the growth rate can 
be rewritten in the form  
\be 
\gamma = \lim_{N \to \infty} {1 \over \pi N} \sum_{k=1}^N 
\log (2 H_k) = \lim_{N \to \infty} {1 \over N} 
\sum_{k=1}^N \gamma_k \, = \gamma_\infty \, , 
\ee 
in agreement with Theorem 5. 

\subsection{Width of Stable and Unstable Zones} 

In the plane of parameters (e.g., Figure \ref{fig:qlplane}), the width
of the stable and unstable zones can be found for the delta function
limit.  In this case, the leading edge of the zone of stability is
given by the condition
\be
\theta = \sqrt{\af} \pi = n \pi \, , 
\ee 
where $n$ is an integer that can be used to label the zone in
question.  The beginning of the next unstable zone is given by the
condition $|h| = 1$.  In the limit of large $q \gg 1$, the width of
the stable regime is narrow, and the boundary will fall at $\theta = n
\pi + \varphi$, where $\varphi$ is small. In particular, $\varphi$
will be smaller than $\pi/2$, so that the angle $\theta$ will lie in
either the first or third quadrant, which in turn implies that $\sin
\theta$ and $\cos \theta$ have the same sign. As a result, the
condition at the boundary takes the form
\be 
{ q \over 2 \sqrt{\af} } = {1 + \cos \varphi \over \sin \varphi}  
\approx {2 \over \varphi} \, . 
\ee 
If we solve this expression for $\varphi$ and use the definition 
$\varphi = \theta - n \pi$, we can solve for the value of $\af$ 
at the boundary of the zone, i.e., 
\be 
\af \approx {n^2 \over (1 - 4/ q \pi)^2} \approx n^2 
\Bigl[ 1 + {8 \over q \pi} + {\cal O} ( q^{-2} ) \Bigr] \, . 
\ee
The width of the stable zone can then be expressed in the form 
\be 
\Delta \af = {8 n^2 \over \pi q} \, . 
\ee
For any finite $q$, there exists a zone number $n$ such that 
$n^2 > q$ and the width of the zone becomes wide. In the limit 
$q \to \infty$, the zones are narrow for all finite $n$. 

Note that when the forcing strength $q_k$ varies from cycle to cycle, 
we can define the expectation value of the zone widths, 
\be 
\left\langle \Delta \lambda \right\rangle = {8 n^2 \over \pi} 
\left\langle {1 \over q_k} \right\rangle \, . 
\ee
This expectation value exists under the same conditions required 
for Theorem 5 to be valid. 

\subsection{Variations in ($\af_k,q_k)$ and Connection to the General Case} 

As outlined earlier, the growth rates $\Delta \gamma$ depend on the
ratios of the principal solutions, rather than on the input parameters
$(\af_k, q_k)$ that appear in the original differential equation
(\ref{eq:basic}). Since we have analytic expressions for the principal
solutions in the delta function limit, we can study the relationship
between the distributions of the fundamental parameters $(\af_k, q_k)$
and the distribution of the composite variable $\xi = \log (x_k/x_j)$
that appears in the Theorems of this paper.

As a starting point, we first consider the limiting case where $q_k
\to \infty$ and the parameter $\af_k$ is allowed to vary. We also
focus the discussion on the correction $\Delta \gamma$ to the growth
rate, which depends on the ratios $x_k$.  In this limit, using 
equation (\ref{eq:elements}), we see the variables
$x_k$ reduce to the simple form
\be 
x_k = {\pi \over \theta_k} {\sin\theta_k \over 1 + \cos\theta_k} \, , 
\label{eq:xkinf} 
\ee
where $\theta_k \equiv \sqrt{\lambda} \pi$. In this case the
distribution of $\xi = \log (x_k/x_j)$ depends only on the
distribution of the angles $\theta_k$, which is equivalent to the
distribution of $\af_k$.  Since the $x_j$ and $x_k$ are drawn
independently from the same distribution (of $\theta_k$), the variance
of the composite variable $\sigma_0^2 = 2 \sigma_x^2$, where
$\sigma_x^2$ is the variance of $\log x_k$.

As a benchmark case, we consider the distribution of $\theta$ to be
uniformly distributed over the interval $[0,2\pi]$. For this example,
\be
\sigma_x^2 = \int_0^{2\pi} {d\theta \over 2 \pi} 
\left[ \log \left( {\pi \over \theta} 
{\sin\theta \over 1 + \cos\theta} \right) \right]^2 - 
\left[ \int_0^{2\pi} {d\theta \over 2 \pi} 
\log \left( {\pi \over \theta} {\sin\theta \over 1 + \cos\theta} \right) 
\right]^2 \, . 
\ee
Numerical evaluation indicates that $\sigma_0 \approx 2.159$. Further,
the correction to the growth rate is bounded by $\Delta \gamma \le
\sigma_0^2/(4 \pi) \approx 0.371$ and is expected to be given
approximately by $\Delta \gamma \sim 0.13$. In this limit we expect
the asymptotic growth rate to dominate.  For example, if $q_k\sim
1000$, a typical value for one class of astrophysical orbits [AB],
then $\gamma_\infty \approx 2$, which is an order of magnitude greater
than $\Delta \gamma$. Note that in the limit of large (but finite)
$q_k$, the corrections to equation (\ref{eq:xkinf}) are of order
${\cal O} (1/q_k)$, which will be small, so that the variance
$\sigma_0^2$ of the composite variable $\xi$ will be nearly
independent of the distribution of $q_k$ in this limit.

As another way to illustrate the transformation between the
$(\af_k,q_k)$ and the matrix elements $x_k$, we consider the case of
fixed $\af_k$ and large but finite (and varying) values of $q_k$.  We
are thus confining the parameter space in Figure \ref{fig:qlplane} to
a particular vertical line, which is chosen to be in an unstable band.
We thus define $\theta = \sqrt{\af} \pi$, and the $x_k$ take the form 
\be
x_k = { q_k (\pi/\theta) \sin \theta - 2 \cos\theta \over 
q_k (1 + \cos\theta)/2 + (\theta/\pi) \sin\theta} \, . 
\ee 
For purposes of illustration, we can make a further simplification by
taking $\theta$ to have a particular value; for example, if $\theta =
\pi/2$, the $x_k$ are given by  
\be 
x_k = {2 q_k \over q_k + 1} \, .
\ee
For this case, the relevant composite variable $\xi$ is given by 
\be
\xi = \log \left[ {q_k \over q_j} {q_j + 1 \over q_k + 1} \right] \, , 
\ee
where $q_j$ and $q_k$ are the values for two successive cycles.  In
the limit of large $q_j, q_k \gg 1$, the composite variable takes the
approximate form $\xi \approx (q_k - q_j)/(q_k q_j)$ which illustrates
the relationship between the original variables (only the $q_k$ in
this example) and the $x_k$, or the composite variable $\xi$, that
appear in the growth rates.

Before leaving this section, we note that the more general case of
Hill's equation with a square barrier of finite width can also be
solved analytically (e.g., let $\qhat(t) = 1/w$ for a finite time
interval of width $\Delta t = w$, with $\qhat (t)$ = 0 otherwise). 
For this case, in the limit of large $q_k$, the solution for $h_k$ 
takes the form 
\be
|h_k| \propto \sin (w q_k)^{1/2} 
\left( {q_k \over w \af_k} \right)^{1/2} \, . 
\ee 
In the limit of large but finite $q_k$ and vanishing width $w \to 0$,
we recover the result from the delta function limit, i.e., the
dependence on the width $w$ drops out and $|h_k| \propto q_k$. In the
limit of finite $w$ and large $q_k$ [specifically, when $(w q_k) \ll
1$ does {\it not} hold], then $|h_k| \propto \sqrt{q_k}$. This example
vindicates our expectation that large $q_k$ should lead to large
$h_k$, but the dependence depends on the shape of the barrier
$\qhat(t)$. An interesting problem for further study is to place
constraints on the behavior of the matrix elements $h_k$ (and $g_k$)
as a function of the forcing strengths $q_k$ for general $\qhat(t)$.

\section{DISCUSSION AND CONCLUSION} 

This paper has considered Hill's equation with forcing strengths and
oscillation parameters that vary from cycle to cycle. We denote such
cases as random Hill's equations.  Our first result is that Hill-like
equations where the period is not constant, but rather varies from
cycle to cycle, can be reduced to a random Hill's equation (Theorem
1). The rest of the paper thus focuses on random Hill's equations,
specifically, general equations in the unstable limit (\S 3) and the
particular cases of the delta function limit (\S 4), where the
solutions can be determined in terms of elementary functions.

For a general Hill's equation in the limit of a large forcing
parameter, we have found general results governing instability. In all
cases, the growth rates depend on the distribution of values for the
elements of the transition matrix that maps the solution for one cycle
onto the next. The relevant composite variable $\xi$ is determined by
the principal solutions via the relation $\xi = \log [y_{1k} (\pi)
{\dot y}_{1j} (\pi) / {\dot y}_{1k} (\pi) y_{1j} (\pi) ]$, where $k$
and $j$ denote two successive cycles; our results are then presented
in terms of the variance $\sigma_0$ of the distribution of $\xi$. The
growth rate can be separated into two parts, the asymptotic growth
rate $\gamma_\infty$ that would result if each cycle grew at the rate
appropriate for an ordinary Hill's equation, and the correction term
$\Delta \gamma$ that results from matching the solutions across
cycles. The asymptotic growth rate $\gamma_\infty$ is determined by
the appropriate average of the growth rates for individual cycles (see
eqs. [\ref{eq:gaminf}] and [\ref{eq:gamtwo}]).  In contrast, the
correction term $\Delta \gamma$ results from a type of random walk
behavior and depends on the variance of the distribution of the
composite variable $\xi$ defined above.

For the case of purely positive matrix elements, the correction term
$\Delta \gamma$ has a simple form (Theorem 2), and is positive
semi-definite and bounded from above and below.  In the limit of small
variance, the correction term $\Delta \gamma \propto \sigma_0^2$,
whereas in the limit of large variance, $\Delta \gamma \propto
\sigma_0$. For all $\sigma_0$, the correction term to the growth rate
is bounded by $\Delta \gamma \le \sigma_0^2/4\pi$ (Theorem 3). A
sharper bound could be obtained in the future.

For the case of matrix elements with varying signs, we have found the
growth rate of instability (Theorem 4), where the results depend on
the probability $p$ of the matrix elements having a positive sign. In
the limit of small variance, the correction term $\Delta \gamma$ is
always negative and approaches the form $\Delta \gamma \propto \log
\sigma_0$ (unless $p$ = 1, where $\Delta \gamma \to 0$ in this
limit). As a result, the total growth rate $\gamma = \gamma_\infty$ +
$\Delta \gamma$ will always be negative -- and hence the system will
be stable -- for sufficiently small variance $\sigma_0$ and any
admixture of mixed signs.  In the opposite limit of large variance,
the growth rate for mixed signs and that for purely positive signs
converge, with both cases approaching the form $\Delta \gamma \propto
\sigma$; the difference between the growth rates for the two cases
decreases as $\Delta (\Delta \gamma) \propto 1/\sigma_0$.

For the delta function limit, we can find the solution explicitly for
each cycle, and thus analytically define the matrix elements of the
discrete map that develops the solution (eq. [\ref{eq:matrixdef}] and
[\ref{eq:elements}]). For the case in which only the forcing strength
varies, the growth rate of the general solution approaches the
asymptotic growth rate (eq. [\ref{eq:gaminf}]), which represents the
growth the solution would have if every cycle grows at the rate
appropriate for a standard (non-stochastic) Hill's equation. We have
calculated the widths of the stable and unstable zones for Hill's
equation in the limit of delta function forcing and large growth
rates, which represents a specific case of the results presented in
[WK], where this specific case includes random forcing terms.
Finally, we have used the analytic solutions for the delta function
limit to illustrate the transformation between the original variables
$(\af_k, q_k)$ that appear in Hill's equation and the variables $x_k$
that determine the growth rates (\S 4.5).

Although this paper takes a step forward in our understanding of
Hill's equation (in particular, generalizing it to include random
forcing terms) and the multiplication of random matrices (of the
particular form motivated by Hill's equation), additional work along
these lines can be carried out. The analysis presented herein works
primarily in the limit of large $q_k$, where the solutions are highly
unstable, although we have bounded the errors incurred by working in
this limit.  Nonetheless, the case in which some cycles have stable
solutions, while others have unstable solutions, should be considered
in greater detail.  This paper presents bounds on the correction term
$\Delta \gamma$ to the growth rate, but a sharper bound could be
found.  In the treatment of this paper, we considered the probability
distribution of the composite variable $\xi = \log (x_k/x_j)$ to be
symmetric, which implies that $x_k$ and $x_j$ are independently drawn
from their distribution. In future work, correlations between
successive cycles can be considered and would lead to asymmetric
probability distributions.  Most of the results of this paper are
presented in terms of the distributions of the composite variables
$x_k$, rather than the original parameters that appear in Hill's
equation; the transformation between the distributions of the $(\af_k,
q_k)$ and the $x_k$ thus represents another interesting problem for
future study. Another case of interest we intend to consider is the
case where $\hat{Q}(t)$ takes the form a finite Fourier series.
Finally, the relationship between solutions to random Hill's equations
and the multiplication of random matrices should be explored in
greater generality.

Random Hill's equations, and the properties of their solutions, have a
wide variety of applications. The original motivation for this work
was a class of orbit problems in astrophysics. In that context, many
astrophysical systems --- young embedded star clusters, galactic
bulges, and dark matter halos --- are essentially triaxial extended
mass distributions. Orbits within these mass distributions are often
chaotic; further, when motion is initially confined to a plane, the
equation of motion for the perpendicular direction is described by a
random Hill's equation. The instability explored here thus determines
how quickly an orbiting body will explore the perpendicular
direction. For example, this class of behavior occurs in young
embedded star clusters, which begin in highly flattened configurations
but quickly become rounder, in part due to the instability described
here. Dark matter halos are found (numerically) to display nearly
universal forms for their density distributions [NF, BE], but an {\it
a priori} explanation for this form remains lacking. Since the orbits
of dark matter particles will be subject to the instability studied
herein, random Hill's equations must play a role in the explanation.
As yet another example, galactic bulges often harbor super-massive
black holes at their centers; the resulting stellar orbits, including
the instability considered here, play a role in feeding stars into the
central black hole. Finally, we note that in addition to astrophysical
applications, random Hill's equations are likely to arise in a number
of other settings.

\bigskip 
\centerline{\bf Acknowledgments} 

We would like to thank Charlie Doering, Gus Evrard, Divakar Vishwanath
and Michael Weinstein for many useful conversations, and research
students Michael Busha, Suzanne Butler, Jeff Druce, Jake Ketchum, and
Eva Proszkow for performing numerical calculations that guided the
initial formulation of this project. We also thank an anonymous
referee for many useful comments and criticisms that improved the
paper. This work was supported at the University of Michigan by the
Michigan Center for Theoretical Physics; by NASA through the Spitzer
Space Telescope Theoretical Research Program; and by NSF through
grants CMS-0408542 and DMS-604307. Some of this work was completed at
the Kavli Institute for Theoretical Physics, at U. C. Santa Barbara,
and was supported in part by the National Science Foundation under
Grant No. PHY05-51164.

\renewcommand{\theequation}{A\arabic{equation}}
\setcounter{equation}{0}  
\section*{Appendix A: Astrophysical Motivation} 

This Appendix outlines the original astrophysical problem that
motivated this study of Hill's equation with random forcing. In the
the initial setting, the goal was to understand orbits in potentials
resulting from a density profile of the form
\be 
\rho = \, \rho_0 \, {f(m) \over m} \, , 
\label{eq:rhogen} 
\ee 
where $\rho_0$ is a density scale. This form arises in many different
astrophysical contexts, including dark matter halos, galactic bulges,
and young embedded star clusters.  The density field is constant on
ellipsoids and the variable $m$ has a triaxial form
\be
m^2 = {x^2 \over a^2} + {y^2 \over b^2} + {z^2 \over c^2} \, , 
\label{eq:mdef} 
\ee 
where, without loss of generality, $a > b > c > 0$. 
The radial coordinate $\xi$ is given by $\xi^2 = x^2 + y^2 + z^2$. 
The function $f(m)$ is assumed to approach unity as $m \to 0$ so that
the density profile approaches the form $\rho \sim 1/m$.  For this
inner limit, one can find an analytic form for both the potential and
the force terms [AB]. For purposes of illustration, we write the force
terms for the three spatial directions in the form 
\be
{\cal F}_x = - {2 x \over F(a)} 
\ln \Bigg| {2 F(a) \sqrt{\Gamma} + 2 \Gamma - \Lambda a^2 \over 
a^2 \bigl[ 2 F(a) \xi + \Lambda  - 2 a^2 \xi^2 \bigr] }
\Bigg| \, , 
\label{eq:xforce} 
\ee
\be
{\cal F}_y = - {2 y \over |F(b)|} 
\Bigl[ \sin^{-1} \Bigl( {\Lambda - 2 b^2 \xi^2 \over 
\sqrt{\Lambda^2 - 4 \xi^2 \Gamma} } \Bigr) - \sin^{-1} \Bigl( 
{2 \Gamma/b^2 - \Lambda \over \sqrt{\Lambda^2 - 4 \xi^2 \Gamma} } 
\Bigr) \Bigr] \, , 
\label{eq:yforce} 
\ee
\be 
{\cal F}_z = - {2 z \over F(c)} 
\ln \Bigg| {2 F(c) \sqrt{\Gamma} + 2 \Gamma - \Lambda c^2 \over 
c^2 \bigl[ 2 F(c) \xi + \Lambda  - 2 c^2 \xi^2 \bigr] }
\Bigg| \, . 
\label{eq:zforce} 
\ee
The coefficients in the numerators are given by the following
quadratic functions of the coordinates:  
\be 
\Lambda \equiv (b^2+c^2) x^2 + (a^2+c^2) y^2 + (a^2+b^2) z^2 
\qquad {\rm and} \qquad 
\Gamma \equiv b^2 c^2 x^2 + a^2 c^2 y^2 + a^2 b^2 z^2 \, , 
\label{eq:gamdef} 
\ee 
and the remaining function $F$ is defined by 
\be 
F(\alpha) \equiv \big[ \xi^2 \alpha^4 - \Lambda \alpha^2  
+ \Gamma \bigr]^{1/2} \, . 
\label{eq:fdef} 
\ee 
Equations (\ref{eq:xforce} -- \ref{eq:fdef}) define the force terms 
that determine the orbital motion of a test particle moving in the 
potential under consideration (i.e., that resulting from a triaxial 
density distribution of the form [\ref{eq:rhogen}]). The work of [AB] 
shows that when the orbit begins in any of the three principal planes, 
the motion is (usually) highly unstable to perturbations in the 
perpendicular direction. For example, for an orbit initially confined 
to the $x-z$ plane, the amplitude of the $y$ coordinate will (usually) 
grow exponentially with time. In the limit of small $y$, the equation 
of motion for the perpendicular coordinate simplifies to the form  
\be 
{d^2 y \over dt^2} + \omega_y^2 y = 0 \qquad {\rm where} \qquad 
\omega_y^2 = { 4/b \over \sqrt{c^2 x^2 + a^2 z^2} + b \sqrt{x^2 + z^2} } \ . 
\label{eq:omegay} 
\ee 
Here, the time evolution of the coordinates $(x,z)$ is determined by
the orbit in the original $x-z$ plane. Since the orbital motion is
nearly periodic, the $(x,z)$ dependence of $\omega_y^2$ represents a
periodic forcing term. The forcing strengths, and hence the parameters
$q_k$ appearing in Hill's equation (\ref{eq:basic}), are thus
determined by the inner turning points of the orbit (with appropriate
weighting from the axis parameters $[a,b,c]$).  Further, since the
orbit in the initial plane often exhibits chaotic behavior, the
distance of closest approach of the orbit, and hence the strength
$q_k$ of the forcing, varies from cycle to cycle. The orbit also has
outer turning points, which provide a minimum value of $\omega_y^2$,
which defines the unforced oscillation frequency $\af_k$ appearing 
in Hill's equation (\ref{eq:basic}). As a result, the equation of
motion (\ref{eq:omegay}) for the perpendicular coordinate has the form
of Hill's equation, where the period, the forcing strength, and the
oscillation frequency generally vary from cycle to cycle.

\renewcommand{\theequation}{B\arabic{equation}}
\setcounter{equation}{0}  
\section*{Appendix B: Growth Rate for an Ancillary Matrix} 

In this Appendix, we separate the transformation matrix for the
general case (not in the limit of small variance) and find the growth
rate for one of the matrices. We include this result because examples
where one can explicitly find the growth rates (Lyapunov exponents)
for random matrices are rare.  Specifically, the transition matrix can
be written in the form given by equation (\ref{eq:cksum}), where the
second term in the sum has the form
\be 
s_k (x_k - 1) {\bf B}_k \qquad {\rm where} \qquad 
{\bf B}_k = \left[ \matrix{0 & 1 \cr -1/x_k & 0 } \right] \, . 
\ee 
Note that any pair of matrices ${\bf A}_k$ with opposite
signs will vanish, and so will all subsequent products.

\noindent 
The products of the second term (the matrices ${\bf B}_k$ along with 
the leading factor) have a well-defined growth rate: 

\noindent
{\bf Proposition 3:} {\it The growth rate of matrix multiplication for the
matrix ${\bf M}_k = (x_k - 1) {\bf B}_k$ is given by} 
\be 
\gamma_B = \lim_{N \to \infty} {1 \over 2 \pi N } \left\{ 
\sum_{k=1}^N \log | x_k - 1 | + \sum_{j=1}^N \log | 1/x_j - 1 | \right\} \, . 
\label{eq:asymp} 
\ee 

\noindent
{\bf Proof}: The products of the matrices ${\bf B}_k$ follow 
cycles as shown by the first three nontrivial cases: 
\be 
{\bf B}_2 {\bf B}_1 = \left[ \matrix{-1/x_1 & 0 \cr 0  & -1/x_2 } \right] \, , 
\qquad {\bf B}_3 {\bf B}_2 {\bf B}_1 = 
\left[ \matrix{0 & -1/x_2 \cr 1/(x_1 x_3) & 0 } \right] \, , 
\ee 
and 
\be 
{\bf B}_4 {\bf B}_3 {\bf B}_2 {\bf B}_1 = 
\left[ \matrix{1/(x_1 x_3) & 0 \cr 0 & 1/(x_2 x_4) } \right] \, . 
\ee
Thus, the even products of the matrices are diagonal matrices, whereas
the odd products produce matrices with only off-diagonal elements.
As a result, the product matrix will approach the form
\be 
{\bf M}^{(N)} \sim \left( \prod_{k=1}^N (x_k - 1) \right)
\left[ \matrix{\podd & 0 \cr 0 & \peven } \right] 
\qquad {\rm or} \qquad 
{\bf M}^{(N)} \sim \left( \prod_{k=1}^N (x_k - 1) \right)
\left[ \matrix{0 & - \peven \cr \podd & 0 } \right] \, , 
\ee
where we have defined 
\be 
\podd \equiv \prod_{k=1, odd}^N {1 \over x_k} 
\qquad {\rm and} \qquad 
\peven \equiv \prod_{k=2, even}^N {1 \over x_k} \, . 
\ee 
For $N$ even (odd), the eigenvalues are $\Lambda = \peven, \podd$
($\Lambda = \pm i \sqrt{\peven \podd}$). Since $|\peven| = |\podd|$ in
the limit $N \to \infty$, the eigenvalues (and hence the growth rates)
have the same magnitudes in either case.  To compute the growth rate
$\gamma_B$, we need to account for the fact that only half of the
factors (either the even or odd terms) appear in the products $\podd$
and $\peven$. After some rearrangement, we obtain equation
(\ref{eq:asymp}). $\thmbox$

\end{document}